\documentclass[]{JHEP3} 

\usepackage{epsfig,multicol,bbm}
\usepackage[sort&compress,numbers]{natbib}

\newcommand\fverb{\setbox\fverbbox=\hbox\bgroup\verb}
\newcommand\fverbdo{\egroup\medskip\noindent%
			\fbox{\unhbox\fverbbox}\ }
\newcommand\fverbit{\egroup\item[\fbox{\unhbox\fverbbox}]}
\newbox\fverbbox

\usepackage{graphicx}
\def\beq{\begin{equation}}
\def\eeq{\end{equation}}
\def\cG{r_\Gamma}

\def\prop#1{{\cal P}_{#1}}

\def\I33m{\mathrm{I}_3^{3{\mathrm m}}}
\def\nn{\nonumber}

\def\be{\begin{equation}}
\def\ee{\end{equation}}
\def\bea{\begin{eqnarray}}
\def\eea{\end{eqnarray}}

\def\e{\epsilon}

\def\spa#1.#2{\left\langle#1\,#2\right\rangle}
\def\spb#1.#2{\left[#1\,#2\right]}
\def\lor#1.#2{\left(#1\,#2\right)}
\def\sand#1.#2.#3{%
\left\langle\smash{#1}{\vphantom1}^{-}\right|{#2}%
\left|\smash{#3}{\vphantom1}^{-}\right\rangle}
\def\sandp#1.#2.#3{%
\left\langle\smash{#1}{\vphantom1}^{-}\right|{#2}%
\left|\smash{#3}{\vphantom1}^{+}\right\rangle}
\def\sandpp#1.#2.#3{%
\left\langle\smash{#1}{\vphantom1}^{+}\right|{#2}%
\left|\smash{#3}{\vphantom1}^{+}\right\rangle}
\def\sandpm#1.#2.#3{%
\left\langle\smash{#1}{\vphantom1}^{+}\right|{#2}%
\left|\smash{#3}{\vphantom1}^{-}\right\rangle}
\def\sandmp#1.#2.#3{%
\left\langle\smash{#1}{\vphantom1}^{-}\right|{#2}%
\left|\smash{#3}{\vphantom1}^{+}\right\rangle}
\def\spab#1.#2.#3{\langle#1|#2|#3]}
\def\spba#1.#2.#3{[#1|#2|#3\rangle}

\def\met{\ifmmode%
\setbox0=\hbox{$E_t$}%
\setbox1=\hbox to\wd0{\hss$/$\hss}\else%
\setbox0=\hbox{E_t}%
\setbox1=\hbox to\wd0{\hss/\hss}\fi%
E_t\hskip-\wd0\box1 }

\title{Gluon-gluon contributions to $W^+ W^-$ production and
Higgs interference effects}

\author{
    John M. Campbell, R. Keith Ellis and Ciaran Williams
    \\
    Fermilab, Batavia, IL 60510, USA
    \\
    E-mails: 
    {\tt johnmc@fnal.gov}, 
    {\tt ellis@fnal.gov}, 
    {\tt ciaran@fnal.gov}.}

\received{\today} 		
\accepted{\today}		

\preprint{
FERMILAB-PUB-11-340-T}

\abstract{In this paper we complete our re-assessment of 
the production of $W$ boson pairs at the LHC, by calculating analytic results for the 
$gg \to W^+W^- \to \nu l^+ l^- \bar{\nu}$ 
process including the effect of massive
quarks circulating in the loop.
Together with the one-loop amplitudes containing the first two generations of massless quarks propagating
in the loop, these diagrams can give a significant contribution with a large flux of gluons.
One of the component parts of this
calculation is the production of a standard model Higgs boson, 
$gg \to H$ and its subsequent decay, $H \to W^+(\to \nu l^+) W^-(\to l^- \bar{\nu})$. 
We will quantify the importance of the interference between the Higgs boson production process 
and the gluon-induced continuum production
in the context of searches for the Higgs boson at the Tevatron and the LHC.
For instance, for $m_H < 140$~GeV the effect of the interference
typically results in around a $10\%$ reduction in the expected number of Higgs signal events.
The majority of this interference is due to non-resonant contributions.
Therefore cuts on the transverse mass such as those currently used
by the ATLAS collaboration reduce the destructive interference to
about a $1\%$ effect. We advocate that a cut on the maximum
transverse mass be used in future Higgs searches in this channel.}


\begin{document} 


\section{Introduction}
The search for the Standard Model (SM) Higgs boson is entering the closing stages. With the LHC acquiring data at an impressive rate, evidence for the existence of the Higgs boson or its 
exclusion over the mass range $115-600$ GeV should be expected by the end of 2012. In addition, ongoing studies at the Tevatron with around 10 $\rm{fb}^{-1}$ will
provide additional independent exclusion regions over a wide range of potential Higgs masses. In summary, the combined results from both hadron colliders can be expected to 
tightly constrain the Higgs boson within the course of the coming year. Winter 2010 results from the various collaborations can be found in refs.~\cite{Chatrchyan:2011tz,Aaltonen:2011gs,Aad:2011qi}. 

An important Higgs search channel over the range $120-200$ GeV is the process $gg\rightarrow H \rightarrow  WW \rightarrow \nu \ell\ell'\nu'$. This proceeds 
through gluon fusion via a top (or bottom) loop with the subsequent production of two $W$ bosons that then decay leptonically. In setting a limit on this channel 
one needs to possess precise predictions for the cross section and kinematic distributions. Thankfully, as a result of much theoretical work the (fully differential) production cross section 
for a Higgs boson through gluon fusion is known at NNLO (see for example refs.~\cite{Anastasiou:2002yz,Ravindran:2003um,Anastasiou:2004xq,Anastasiou:2007mz,Catani:2007vq,Grazzini:2008tf,Anastasiou:2011pi}).
By including the NNLO corrections one typically finds scale uncertainties on inclusive quantities (which are used to quantify theoretical uncertainty by the collaborations) of $\mathcal{O}(10-20\%)$~\cite{Dittmaier:2011ti}.
Quoting an error of this order 
requires that all contributions to the cross section that could possibly change it by $\mathcal{O}(10\%)$ have been calculated, and included in the theoretical predictions used. 
To correctly exclude the Higgs then, one must confirm that no remaining $\mathcal{O}(10\%)$ effects are neglected in the theoretical prediction. 

One potentially large contribution to the cross section is the interference between the Higgs signal and SM continuum production of $WW$ pairs. Power counting 
in $\alpha_S$ indicates that this contribution is the same order as the LO Higgs signal cross section. 
Typically these interference terms are neglected in the theoretical calculations used 
by the collaborations. The aim of this paper is to fully quantify this interference under a variety of experimental cuts.  In order to do this we calculate 
analytic results for the $gg \to WW \to \nu l^+ l^- \bar{\nu}$ process
including the effect of massive  quarks circulating in the loop. 
The interference has been studied in ref.~\cite{Binoth:2006mf}, in which the focus was on a $14$ TeV LHC, and on the effects on the total cross section. 
In this study we will re-address the issue in the context of current
Higgs boson searches. 
The interference effects associated with Higgs production and decay to $ZZ$ have been studied some time ago~\cite{Glover:1988fe,Glover:1988rg}. In these papers 
the primary focus was at SSC operating energies and large Higgs masses. Similar studies have also been performed for the case of a light Higgs boson
decaying into di-photons at the LHC~\cite{Dixon:2003yb} and the case of Higgs production at a photon collider with subsequent decay
into bottom quark pairs~\cite{Dixon:2008xc}.

The process $q\bar{q} \to WW$ was first calculated in the Born approximation in ref.~\cite{Brown:1978mq}, with strong corrections later
computed in refs.~\cite{Ohnemus:1991kk,Frixione:1993yp,Ohnemus:1994ff,Dixon:1998py}. Phenomenological NLO results for the Tevatron
and the LHC have been presented in refs.~\cite{Campbell:1999ah,Dixon:1999di,Campbell:2011bn}. The gluon-initiated contribution
$gg\rightarrow WW$ was first calculated in refs.~\cite{Dicus:1987dj,Glover:1988fe} and later improved to account for  
off-shell effects of the vector bosons and their subsequent decays in ref.~\cite{Binoth:2005ua}. The effect of massive quarks circulating 
in the loop was later assessed in ref.~\cite{Binoth:2006mf}. Combined results, including both the $q\bar q$ and $gg$ initiated contributions --
for loops of massless quarks only -- were recently presented in ref.~\cite{Campbell:2011bn}. In this paper we shall use modern methods to provide a calculation
of the massive quark contributions that is equivalent to the calculation in ref.~\cite{Binoth:2006mf} and update the
results of ref.~\cite{Campbell:2011bn} accordingly.

The plan of this paper is as follows. In section 2 we shall describe the calculation of the amplitude $gg \to WW$ where a massive top quark (and, where numerically
significant, a massive bottom quark) circulates in the loop. Sections 3, 4 and 5 will present phenomenological results using the implementation of
these amplitudes into the parton level Monte Carlo program MCFM. Section 3 will address the size of the corrections to the $WW$ process
resulting from the inclusion of massive quarks. In sections 4 and 5 we shall discuss the interference contributions at the LHC and Tevatron respectively.
In particular we will discuss these effects in the region where cuts are applied to enhance the signal to background ratio for the detection
of the Standard Model Higgs boson. In section 6 we discuss the impact of the interference in regards to the NNLO predictions for the total cross section. Finally,
in section 7 we draw our conclusions.

\section{Calculation}
In reference~\cite{Campbell:2011bn} 
we showed that the calculation of $gg \to WW$ with massless quarks
circulating in the loop can be extracted from Ref.~\cite{Bern:1997sc}.
Specifically, we find that the contribution from a single generation of
massless quarks in the loop is given by,
\begin{eqnarray}
&& {\cal A}_{6}^{1\rm -loop}\left(1_g^{h_1},2_g^{h_2},3_{\nu_\ell}^-,4_{\overline{\ell}}^+,5_{\ell^\prime}^-,6_{\overline{\nu}_{\ell^\prime}}^+ \right)
\nn \\
&=& \delta^{a_1 a_2} \left(\frac{g_w^4 g_s^2}{16\pi^2}\right) \prop{W}(s_{34}) \prop{W}(s_{56})
 \, \left[ -\frac{1}{c_\Gamma} A_{6;4}^v \left(4_q^+, 3_{\bar q}^-, 1_g^{h_1}, 2_g^{h_2}; 5_{\overline e}^-, 6_e^+ \right) \right] \;.
\label{eq:masslessamp}
\end{eqnarray}
The helicities and colour labels of the two gluons are $h_1,h_2$ and
$a_1,a_2$ respectively and the propagator factor is,
\begin{equation}
\prop{W}(s) = {s \over s - m_W^2 + i \,\Gamma_W \, m_W}\,,
\label{eq:propdef}
\end{equation}
where $m_W$ and $\Gamma_W$ are the mass and width of the $W$ boson.  
The amplitude $A_{6;4}^v$ and constant $c_\Gamma$ are defined in
Sections 2, 6 and 11 of ref.~\cite{Bern:1997sc}.  The particle labelling
on the right hand side of this equation is as written in
ref.~\cite{Bern:1997sc}. For our purposes in Eq.~(\ref{eq:masslessamp}) we make the identification,
($q \to \overline{\ell}$, $\bar q \to \nu_\ell$, $\bar e \to \ell^\prime$ and 
$e \to {\overline{\nu}_{\ell^\prime}}$) in going from right to left.

The calculation in Ref.~\cite{Bern:1997sc} was performed using unitarity methods. 
 Our aim in this section is to perform a similar calculation
but including the mass of the top quark circulating in the loop and 
also the mass of the bottom quark where numerically significant.

\subsection{Amplitudes for $gg \to WW$}

\begin{figure}
\begin{center}
\includegraphics[angle=270,scale=0.5]{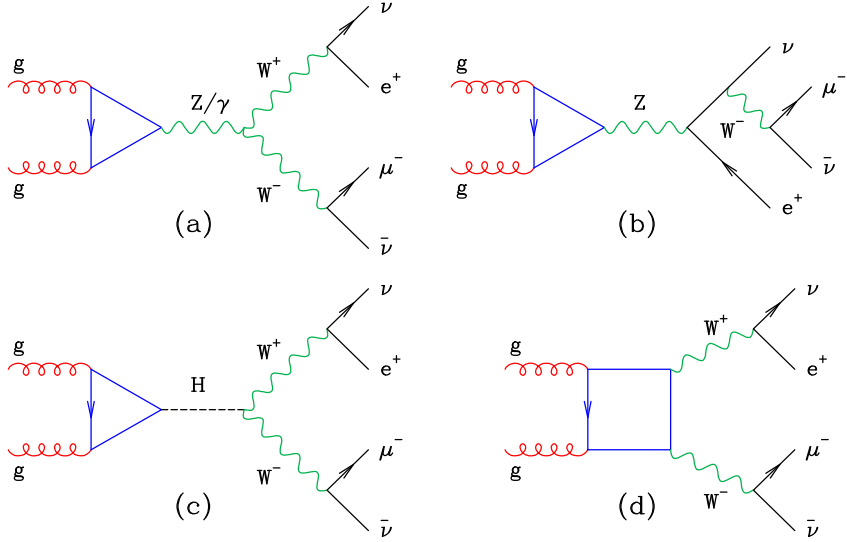}
\end{center}
\caption{Topologies of diagrams that could potentially contribute the process $gg \to \nu_e e^+ \mu^- \bar{\nu}_\mu$.}
\label{topol}
\end{figure}

In this section we present results for the amplitudes relevant for the process,
\begin{equation}
 0 \rightarrow g(p_1) + g(p_2) + \nu_\ell(p_3) + \overline{\ell}(p_4) + \ell^\prime(p_5) + \overline{\nu}_{\ell^\prime}(p_6) \;.
\end{equation}
Topologies of diagrams that could potentially contribute to the leptonic final state
in which we are interested are shown in Fig.~\ref{topol}.
When the vector boson couples to the triangular fermion loop (diagrams (a) and (b))
via a vector coupling, the contribution vanishes via Furry's theorem. When the vector boson couples via
an axial coupling to massless quarks, the contribution of up-type quarks with weak isospin 1/2 and 
down-type quarks with weak isospin -1/2 cancel. In the case of the third generation with a massive top quark 
this cancellation is no longer operative, but it turns out that there is no net contribution~\cite{Binoth:2006mf}
as we shall now illustrate.
If the masses of the appropriate lepton pairs are constrained to have the mass of the $W$, 
the contribution of diagram (a) vanishes and diagram (b) is not present.
This follows because the only non-vanishing contribution of the triangle diagram is
proportional to $p_Z^\mu$ where $p_Z$ is the momentum of the $Z$ boson and $\mu$ is its Lorentz index~\cite{Campbell:2007ev}.
If this condition is relaxed to allow off-shell $W$-bosons,
the contribution of an individual fermion flavour cancels between the doubly resonant diagrams
(a) and the singly resonant diagrams (b). Therefore in both cases the triangle diagrams vanish because of electroweak gauge
invariance. Thus the only triangle loop contribution comes from the
Higgs boson mediated diagram (c).

\begin{figure}
\begin{center}
\includegraphics[angle=270,scale=0.5]{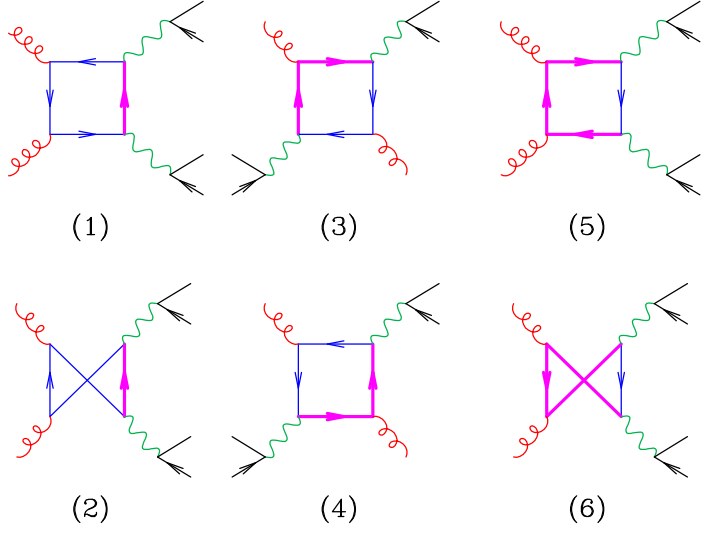}
\end{center}
\caption{Diagrams that contribute to $gg \to WW$ with the third generation of
quarks running in the loop. The top quark is denoted by a heavy (magenta) line
whereas the $b$ bottom quark is shown in blue. The same diagram topologies contribute to
the case where first and second generation quarks run in the loop.}
\label{WWfig}
\end{figure}

We shall therefore separate the calculation of the full amplitude as follows,
\beq
{\cal A}_{\rm full} =
 \delta^{a_1 a_2} \left(\frac{g_w^4 g_s^2}{16\pi^2}\right) \prop{W}(s_{34}) \prop{W}(s_{56})
 \, \left[ 2\,{\cal A}_{\rm massless} + {\cal A}_{\rm massive} + {\cal A}_{\rm Higgs} \right] \;,
\label{eq:ampdecomp}
\eeq
where the first two terms in the square brackets represent the six continuum diagrams shown in Fig.~\ref{WWfig}.
We explicitly separate out the contribution from the first two generations, $(2\,{\cal A}_{\rm massless})$, in
which we consider the quarks in the loop to be massless, and the contribution from the third generation in
which the top quark mass is not neglected, ${\cal A}_{\rm massive}$.
We have also extracted an overall color, coupling and propagator factor that is the same as in the
massless case, so that comparing with Eq.~(\ref{eq:masslessamp}) we immediately see that,
\begin{equation}
{\cal A}_{\rm massless} = -\frac{1}{c_\Gamma} A_{6;4}^v \;.
\end{equation}
The final contribution ${\cal A}_{\rm Higgs}$ originates from two triangle diagrams of the type
shown in Fig.~\ref{topol}(c).

\subsection{Integral basis}
We consider the one-loop amplitude ${\cal A}_{\rm massive}$, corresponding
to third generation quarks circulating in the loop. In our calculation we retain the dependence on the
mass of the top quark but treat the $b$-quark as massless since the effect of a non-zero
value for $m_b$ is at the level of $0.1$\%~\cite{Binoth:2006mf}.

The amplitude can be expanded in terms
of scalar integrals and a rational part as follows,
\beq
{\cal A}_{\rm massive} = \sum_{i=1}^{6} d_i D^{(i)} + \sum_{i=1}^{12} c_i C^{(i)}+ \sum_{i=1}^{6} b_i B^{(i)} + \cal{R} \; .
\label{eq:basicexpansion}
\eeq
The scalar integrals $B^{(i)}$, $C^{(i)}$, $D^{(i)}$ that contribute to this particular amplitude are listed
in Table~\ref{integraldefns}. The
coefficients $b_i$, $c_i$, $d_i$ are amenable to calculation by unitarity cuts; the contribution $\cal{R}$ represents
the remaining non-cut-constructible rational part. We have performed a complete independent numerical check of our
analytic calculation of the coefficients and the rational part using $D$-dimensional unitarity~\cite{Ellis:2008ir}.
No tadpole contributions are found to be present. 
\def\stq{s_{34}}
\def\sud{s_{12}}
\def\scs{s_{56}}
\def\sutq{s_{134}}
\def\sucs{s_{156}}
\def\cI{\cal I}
\begin{table}
\begin{center}
\begin{tabular}{|l|l|l|l|}
\hline
$ D^{(1)} $          & $ I_4( \stq,   0,   0, \scs ; \sutq, \sud ; m^2,   0,   0,   0) $ & $ C^{(1)} $            & $ I_3(   0,   0, \sud ;    0,   0,   0) $  \\ 
$ D^{(2)} $          & $ I_4( \scs,   0,   0, \stq ; \sucs, \sud ; m^2,   0,   0,   0) $ & $ C^{(2)} $            & $ I_3(   0,\sutq, \stq ;    0,   0,m^2) $  \\ 
$ D^{(3)} $          & $ I_4( \scs,   0, \stq,   0 ; \sucs,\sutq ;    0,m^2,m^2,   0) $  & $ C^{(3)} $            & $ I_3(   0, \scs,\sutq ;    0,   0,m^2) $  \\ 
$ D^{(4)} $          & $ I_4( \scs,   0, \stq,   0 ; \sucs,\sutq ; m^2,   0,   0,m^2) $  & $ C^{(4)} $            & $ I_3(   0,\sucs, \scs ;    0,   0,m^2) $  \\ 
$ D^{(5)} $          & $ I_4( \stq,   0,   0, \scs ; \sutq, \sud ;    0,m^2,m^2,m^2) $   & $ C^{(5)} $            & $ I_3(   0, \stq,\sucs ;    0,   0,m^2) $  \\ 
$ D^{(6)} $          & $ I_4( \scs,   0,   0, \stq ; \sucs, \sud ;    0,m^2,m^2,m^2) $   & $ C^{(6)} $            & $ I_3( \sud, \scs, \stq ;    0,   0,m^2) $  \\
\cline{1-2}											 
$ B^{(1)} $            & $ I_2( \sud;    0,   0) $ 				         & $ C^{(7)} $  	  & $ I_3(\sutq,   0, \stq ;	0,m^2,m^2) $  \\
$ B^{(2)} $            & $ I_2( \stq;    0,m^2) $  				         & $ C^{(8)} $  	& $ I_3( \scs,   0,\sutq ;    0,m^2,m^2) $  \\
$ B^{(3)} $            & $ I_2( \scs;    0,m^2) $  				         & $ C^{(9)} $  	& $ I_3(\sucs,   0, \scs ;    0,m^2,m^2) $  \\
$ B^{(4)} $            & $ I_2(\sutq;    0,m^2) $  				         & $ C^{(10)} $ 	& $ I_3( \stq,   0,\sucs ;    0,m^2,m^2) $  \\
$ B^{(5)} $            & $ I_2(\sucs;    0,m^2) $  				         & $ C^{(11)} $ 	& $ I_3( \scs, \sud, \stq ;    0,m^2,m^2) $  \\
$ B^{(6)} $            & $ I_2( \sud; m^2,m^2) $      			      		 & $ C^{(12)} $ 	& $ I_3( \sud,   0,   0 ; m^2,m^2,m^2) $  \\
\hline
\end{tabular}
\end{center}
\caption{Definitions of the scalar integrals that appear in the calculation
of the amplitude ${\cal A}_{\rm massive}$, i.e. continuum production of $gg \to W^+W^-$ through a loop containing at
least one massive particle.}
\label{integraldefns}
\end{table}

The $D$-dimensional scalar integrals themselves are defined as follows,
\begin{eqnarray}
&& I_2(p_1^2;m_1^2,m_2^2)  =
 \frac{\mu^{4-D}}{i \pi^{\frac{D}{2}}\cG}\int d^D l \;
 \frac{1}
{(l^2-m_1^2+i\varepsilon)
((l+q_1)^2-m_2^2+i\varepsilon)}\,,\nn \\
&& I_3(p_1^2,p_2^2,p_3^2;m_1^2,m_2^2,m_3^2)  =
\frac{\mu^{4-D}}{i \pi^{\frac{D}{2}}\cG}
\nn \\
&& \times \int d^D l \;
 \frac{1}
{(l^2-m_1^2+i\varepsilon)
((l+q_1)^2-m_2^2+i\varepsilon)
((l+q_2)^2-m_3^2+i\varepsilon)}\,, \\
&&\nn \\
&&
I_4(p_1^2,p_2^2,p_3^2,p_4^2;s_{12},s_{23};m_1^2,m_2^2,m_3^2,m_4^2)
= 
\frac{\mu^{4-D}}{i \pi^{\frac{D}{2}}\cG}
\nn \\
&&
\times \int d^D l \;
 \frac{1}
{(l^2-m_1^2+i\varepsilon)
((l+q_1)^2-m_2^2+i\varepsilon)
((l+q_2)^2-m_3^2+i\varepsilon)
((l+q_3)^2-m_4^2+i\varepsilon)}\,, \nn
\end{eqnarray}
where $q_n\equiv \sum_{i=1}^n p_i$ and $s_{ij}=
(p_i+p_j)^2$. For the purposes of this paper we take the masses in the
propagators to be real.  Near four dimensions we use $D=4-2 \e$ (and for
clarity the small imaginary part which fixes the analytic
continuations is specified by $+i\,\varepsilon$).  $\mu$ is a scale introduced so that the integrals
preserve their natural dimensions, despite excursions away from $D=4$.
We have removed the overall constant which occurs in $D$-dimensional integrals 
\beq
\cG\equiv\frac{\Gamma^2(1-\e)\Gamma(1+\e)}{\Gamma(1-2\e)} = 
\frac{1}{\Gamma(1-\e)} +{\cal O}(\e^3) =
1-\e \gamma+\e^2\Big[\frac{\gamma^2}{2}-\frac{\pi^2}{12}\Big]
+{\cal O}(\e^3)\,.
\eeq
The final numerical evaluation of the amplitudes uses the QCDLoop library~\cite{Ellis:2007qk} to
provide values for these scalar integrals.

\subsection{Rational part}

In calculating the box diagrams we shall start with the least challenging terms in Fig.~\ref{WWfig} using
the decomposition of Eq.~(\ref{eq:basicexpansion}). The rational terms are completely
determined by the massless calculation which has been already presented. The box diagrams are of rank four, i.e.
they contain at most four powers of the loop momentum in the numerator. However the first mass dependent terms 
contain at least two powers of the mass from numerator factors and are therefore at most of rank two. 
These contributions are
cut-constructible \cite{Bern:1997sc} and do not contribute to the rational part. The rational terms are controlled by the
ultraviolet behavior and thus the presence of  masses in the denominators of the rank 3 and 4 boxes does not
change the rational part. The rational part is therefore identical to the massless case.
This assertion has been checked by direct numerical evaluation.

\subsection{Bubble coefficients}
Turning now to the coefficients of the bubble integrals, we find numerically that the results for the massless case 
can be re-used for the massive case. In the limit that the internal mass $m$ vanishes, we see that the
integrals $B^{(1)}$ and $B^{(6)}$ become degenerate. Therefore the integral $B^{(6)}$ can be dropped from the basis
and we can write the bubble contributions
to the amplitude for the massless calculation as a sum over only 5 terms,
\beq
\sum_{i=1,5} \, b^{(i)}_{\rm massless} \, B^{(i)}_{\rm massless} \;.
\eeq
In this equation the subscript ``massless'' labels the coefficient in the massless calculation and
$B^{(i)}_{\rm massless} = B^{(i)}|_{m=0}$. We find that four of the integral coefficients are
identical to the massless case,
\beq
b^{(i)} = b^{(i)}_{\rm massless} \quad {\rm for}~i=2,3,4,5 \;.
\eeq
Moreover, the sum of the two remaining coefficients is equal to the final massless coefficient,
\beq
b^{(1)} + b^{(6)} =  b^{(1)}_{\rm massless} \;,
\eeq
as a result of the fact that the amplitude does not contain any
poles of ultraviolet origin,
\beq
\sum_{i=1,6} b^{(i)} = 0 \;.
\eeq
We have performed an explicit calculation of the simplest coefficient ($b^{(1)}$) (using the technique described in \cite{Mastrolia:2009dr}) and have hence
determined all six of the bubble coefficients from the known massless bubble coefficients in
Ref.~\cite{Bern:1997sc}.

\subsection{Box coefficients}

We express the amplitudes in terms of spinor products defined as,
\begin{equation}
\spa i.j=\bar{u}_-(p_i) u_+(p_j), \;\;\;
\spb i.j=\bar{u}_+(p_i) u_-(p_j), \;\;\;
\spa i.j \spb j.i = 2 p_i \cdot p_j\,,
\end{equation}
and we further define the spinor sandwich,
\begin{equation}
\spab i.(j+k).l = \spa i.j \spb j.l +\spa i.k \spb k.l \,.
\end{equation}
In order to express the amplitudes compactly we define two symmetry operations,
\begin{eqnarray}
{\rm flip}_1 &:=& \Big\{ (3\leftrightarrow 6),
(4\leftrightarrow 5),\langle \rangle \leftrightarrow []\Big\} \;, \label{eq:flip1def} \\
{\rm flip}_2 &:=& \Big\{ (1\leftrightarrow 2), (3\leftrightarrow 6),
(4\leftrightarrow 5),\langle \rangle \leftrightarrow []\Big\} \;, \label{eq:flip2def} 
\end{eqnarray}
which we will use to relate various coefficients in the amplitude.
Specifically for the box coefficients in Eq.~(\ref{eq:basicexpansion}) related to the
box integrals with the labels given in Table~\ref{integraldefns} we have,
\begin{equation}
d_{2,6}^{++}={ \rm flip}_1~d_{1,5}^{--}, \;\;\; 
d_{2,6}^{--}={ \rm flip}_1~d_{1,5}^{++}, \;\;\; 
d_{2,6}^{-+}={ \rm flip}_1~d_{1,5}^{+-}, \;\;\; 
d_{2,6}^{+-}={ \rm flip}_1~d_{1,5}^{-+} \;.
\end{equation}
Finally the relations between the coefficients for boxes 3 and 4 are,
\begin{equation}
d_{4}^{++}={ \rm flip}_2~d_{3}^{--}, \;\;\;
d_{4}^{--}={ \rm flip}_2~d_{3}^{++}, \;\;\;
d_{4}^{-+}={ \rm flip}_2~d_{3}^{-+}, \;\;\;
d_{4}^{+-}={ \rm flip}_2~d_{3}^{+-} \;.
\end{equation}
Note, in particular, that the ${\rm flip}_2$ symmetry relates
the $(-,+)$ coefficients of boxes 3 and 4 whereas  ${\rm flip}_1$ relates
$d_2^{-+}$ to $d_1^{+-}$.

Given these relations, it is therefore sufficient to compute the coefficients
$d_{1},d_{3}$ and $d_{5}$, which we do using the quadruple cut method \cite{Britto:2004nc}. The simplest coefficients are those of the boxes
containing only a single massive propagator,
\begin{eqnarray}
d_1^{++}&=&0 \;, \nn \\
d_1^{--}&=&0 \;, \nn \\
d_1^{+-}
&=& -\frac{i}{2}\; \frac{s_{12}}{s_{34} s_{56}}
       \frac{\spa1.3 \spb2.6  (m^2-s_{134})^3 }{\spab1.(3+4).2^4} 
( \spa1.3 \spb3.4 \spa5.6 \spb2.6 + m^2 \spa1.5 \spb4.2 )
\nn \\
d_1^{-+} &=&-\frac{i}{2}\; \frac{s_{12}}{s_{34} s_{56}}
\frac{(m^2-s_{134})}{\spab2.(3+4).1^4}
        (m^2 \spb4.1 \spa2.5+\spab2.(1+3).4 \spab5.(3+4).1) \;, \nn \\
&&        (m^2 \spb1.6-\spb5.6 \spab5.(2+6).1)
        (m^2 \spa2.3+\spa3.4 \spab2.(1+3).4) \;.
\label{eq:d1coeff}
\end{eqnarray}
The coefficients of the boxes containing two massive propagators are more
complicated but still relatively compact,
\begin{eqnarray}
d_3^{++}&=&\frac{i}{2}\; \frac{1}{s_{34} s_{56}}
      \frac{\left(s_{134} s_{156}-s_{56} s_{34}+m^2 s_{12}\right) \spa2.3}{s_{12}^2 \spa1.2^2} 
       \left(\spa1.5 \spb5.6 \spb1.2 +\spb1.6 \frac{m^2 s_{12}} {\spab2.(5+6).1}\right) \nn \\
       &&\left(\spa1.5 \spa2.3 \spb3.4 \spb1.2
       - \spa2.5 \spb1.4\frac{m^2 s_{12}}{\spab2.(5+6).1}\right) \;, \nn \\
d_3^{+-}&=&\frac{i}{2}\; \frac{1}{s_{34} s_{56}}
\frac{m^2 \left(s_{134} s_{156}-s_{56} s_{34}+m^2 s_{12}\right) \spb2.6}
 { s_{12}^2 {\spab1.(5+6).2}^2 }
  \left(\spa1.2 \spa4.3 \spb1.4-\spa1.3 \frac{m^2 s_{12}}{\spab1.(5+6).2} \right) \nn \\
       &&\left(\spa1.2 \spb1.4 \spab5.(3+4).2 - \spa1.5 \spb2.4 \frac{m^2 s_{12}}{\spab1.(5+6).2}\right) \;, \nn \\
d_{3}^{--}&=&{ \rm flip}_1~d_{3}^{++} \;, \nn \\
d_{3}^{-+}&=&{ \rm flip}_1~d_{3}^{+-} \;. 
\end{eqnarray}
The coefficients of box 5 are lengthier; interested readers may inspect their form in
the distributed MCFM code.

\subsection{Triangle coefficients}
\begin{figure}
\begin{center}
\includegraphics[scale=0.6,angle=270]{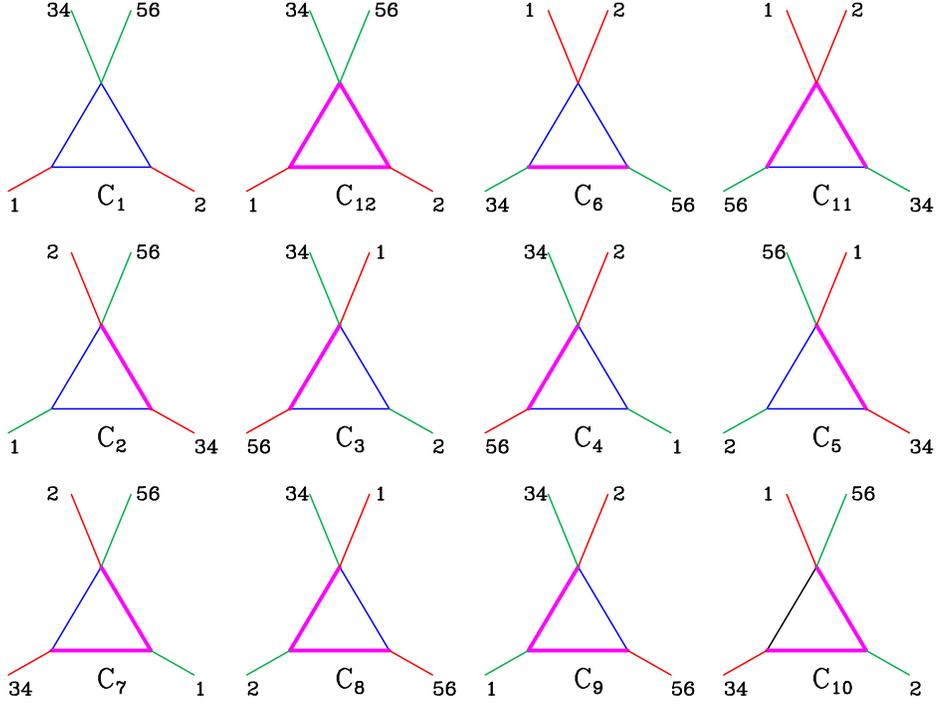}
\end{center}
\caption{Scalar triangle integrals present in the calculation.
\label{fig:tri}}
\end{figure}

Turning now to the coefficients of triangle integrals, we see from Table~\ref{integraldefns} that twelve coefficients are required.
These triangle integrals are depicted in Figure~\ref{fig:tri}.
Since the amplitude that we are calculating is finite, all of the poles in $\epsilon$ must cancel. This cancellation of
infrared poles leads to relations between the triangle coefficients and the singular boxes 1--4.
In particular we find that $c_1$, $c_2$, $c_3$, $c_4$ and $c_5$ can be determined in terms of these box coefficients through
the set of equations,
\begin{eqnarray}
&&\frac{d_1}{s_{134}-m^2}+\frac{d_2}{s_{156}-m^2} + c_1 =0 \;, \\
&&\frac{d_4}{\Delta}+\frac{d_1}{s_{12} (s_{134}-m^2)} + \frac{c_2}{s_{134}-s_{34}} =0 \;, \\
&&\frac{d_3}{\Delta}+\frac{d_1}{s_{12} (s_{134}-m^2)} + \frac{c_3}{s_{134}-s_{56}} =0 \;, \\
&&\frac{d_4}{\Delta}+\frac{d_2}{s_{12} (s_{156}-m^2)} + \frac{c_4}{s_{156}-s_{56}} =0 \;, \\
&&\frac{d_3}{\Delta}+\frac{d_2}{s_{12} (s_{156}-m^2)} + \frac{c_5}{s_{156}-s_{34}} =0 \;.
\end{eqnarray}
These relations involve a kinematic factor $\Delta$ related to boxes 3 and 4, where $\Delta$ is given by, 
\beq
\Delta= (m^2-\sutq)(m^2-\sucs)-(m^2-\stq)(m^2-\scs) \;.
\eeq
Therefore it only remains to determine $c_i$ for $i=6$--$12$. Many of these coefficients are related to one
another by the flip operations defined in Eqs.~(\ref{eq:flip1def},\ref{eq:flip2def}) or simpler symmetries.
In particular we find that the different helicity amplitudes for triangle 6 are related,
\begin{eqnarray}
c_{6}^{--}&=& c_{6}^{++} (1\leftrightarrow 2) \;, \nn \\
c_{6}^{-+}&=& c_{6}^{+-} (1\leftrightarrow 2) \;,
\end{eqnarray} 
whereas the ${\rm flip}_1$ symmetry of Eq.~(\ref{eq:flip1def}) relates the coefficients of triangles
7 and 9,
\begin{eqnarray}
c_{9}^{--}&=& {\rm flip}_1~c_{7}^{++} \;, \nn \\
c_{9}^{++}&=& {\rm flip}_1~c_{7}^{--} \;, \nn \\
c_{9}^{-+}&=& {\rm flip}_1~c_{7}^{+-} \;, \nn \\
c_{9}^{+-}&=& {\rm flip}_1~c_{7}^{-+} \;.
\end{eqnarray} 
The coefficients $c_8$ and $c_{10}$ can be determined from those for $c_7$ and $c_9$ using the
relations,
\begin{eqnarray}
c_{8,10}^{--}&=& c_{9,7}^{--} (1\leftrightarrow 2) \;, \nn \\
c_{8,10}^{++}&=& c_{9,7}^{++} (1\leftrightarrow 2) \;, \nn \\
c_{8,10}^{-+}&=& c_{9,7}^{+-} (1\leftrightarrow 2) \;, \nn \\
c_{8,10}^{+-}&=& c_{9,7}^{-+} (1\leftrightarrow 2) \;.
\end{eqnarray}
Finally, we have the relationships between the different helicity amplitudes for triangle 11,
\begin{eqnarray}
c_{11}^{--}&=& c_{11}^{++} \Big\{ (1\leftrightarrow 2), (3\leftrightarrow 4),
(5\leftrightarrow 6),\langle \rangle \leftrightarrow []\Big\} \;, \nn \\
c_{11}^{-+}&=& c_{11}^{+-} (1\leftrightarrow 2) \;,
\end{eqnarray}
and triangle 12,
\begin{eqnarray}
c_{12}^{--}&=& {\rm flip}_1~c_{12}^{++} \;, \nn \\
c_{12}^{-+}&=& {\rm flip}_1~c_{12}^{+-} \;.
\end{eqnarray} 
We have calculated the coefficients of the remaining unconstrained triangles using the method of 
\cite{Forde:2007mi}.
The results for these triangle coefficients are complicated but may be inspected in
the distributed MCFM code.

\subsection{Calculation of the Higgs mediated diagrams}
The calculation of Higgs boson production via a loop of heavy quarks is well known.
Here we also include the decay of the Higgs boson into a pair of $W$ bosons, with our usual particle
labelling. The contribution to the full amplitude is then,
\begin{equation}
{\cal A}_{\rm Higgs}^{h_1,h_2} =  K^{h_1,h_2} \, \prop{H}{(s_{12})} 
 \sum_{q=t,b} \Bigl[ c_H(m_q) \; I_3(s_{12},0,0; m_q^2,m_q^2,m_q^2) + {\cal R}_H(m_q) \Bigr] \;,
\label{eq:Higgsamp}
\end{equation}
where overall factors of color, couplings and $W$ propagators have
been extracted in Eq.~(\ref{eq:ampdecomp}). Note that here we sum over both top
and bottom quark loops since the effect of including the bottom quark contribution is numerically
important for lower Higgs masses (through its interference with the top quark loops).
Although we have therefore treated the bottom quark mass in an inconsistent manner (finite in the loops
coupling to a Higgs boson, zero otherwise), we have checked that the discrepancy is not phenomenologically
relevant. Although including the mass of the bottom quark can significantly change the total production rate for a light Higgs,
the interference terms that are the focus of this paper have a very small dependence on $m_b$. 
 The dependence on the helicities of the gluons is encoded in the factor $K^{h_1,h_2}$
which is given by,
\begin{eqnarray}
K^{-+} &=& K^{+-} = 0 \;, \nn \\
K^{--} &=& \frac{2 \spa1.2 \spa3.5 \spb6.4}{\spb2.1 s_{34} s_{56}} \;, \nn \\
K^{++} &=& \frac{2 \spb1.2 \spa3.5 \spb6.4}{\spa2.1 s_{34} s_{56}} \;.
\end{eqnarray}
The coefficient of the scalar triangle and the rational contribution are,
\be
c_H(m) = -\frac{im^2}{2} \left(1 - \frac{4m^2}{s_{12}} \right) \;, \qquad
{\cal R}_H(m) = \frac{im^2}{s_{12}} \;.
\ee
Substituting for $I_3(s_{12},0,0; m_q^2,m_q^2,m_q^2)$ in Eq.~(\ref{eq:Higgsamp}) we obtain,
\begin{equation}
{\cal A}_{\rm Higgs}^{h_1,h_2} =  \frac{i K^{h_1,h_2} \, \prop{H}{(s_{12})}}{2}
 \sum_{q=t,b} \, \frac{m_q^2}{s_{12}} \left[  2 + \left(\frac{4m_q^2}{s_{12}}-1\right)F\left(\frac{m_q^2}{s_{12}}\right) \right] \;,
\end{equation}
where the function $F(x)$ is defined by,
\begin{equation}
F(x) = \cases{ \frac{1}{2} \left[ \log \left( \frac{1+\sqrt{1-4x}}{1-\sqrt{1-4x}} \right) - i\pi \right]^2, & $x < 1/4$, \cr
 -2 \left[ \sin^{-1} (1/2\sqrt x) \right]^2, &  $x \ge 1/4$. \cr }
\end{equation}
Up to overall factors that are associated with the decay $H \to WW$, this is the well-known result (see e.g. Ref.~\cite{Ellis:1991qj}).

\section{Effect of massive loops in $gg \to WW$}

\subsection{Parameters}

The results presented in this paper are obtained with the latest version of
the MCFM code (v6.1). We use the default set of electroweak parameters
that assumes the following set of inputs,
\begin{eqnarray}
m_W &=& 80.398~\mbox{GeV}\;, \;\; m_Z=91.1876~\mbox{GeV}\;, \nn \\
\Gamma_W &=& 2.1054~\mbox{GeV}\;, \;\; \Gamma_Z=2.4952~\mbox{GeV} \;, \\
G_F &=& 1.16639 \times 10^{-5} \, \mbox{GeV}^{-2}\;. \nn 
\end{eqnarray}
Using the values of $m_W$, $m_Z$ and $G_F$ as above then determines 
$\alpha_{e.m.}(m_Z)$ and $\sin^2\theta_w$ as outputs, where $\theta_w$ is
the Weinberg angle. We find,
\begin{eqnarray}
\sin^2\theta_w &=& 1 - m_W^2/m_Z^2 = 0.222646 \;, \nn \\
\alpha_{e.m.}(m_Z) &=& \frac{\sqrt2 G_F m_W^2 \sin^2\theta_w}{\pi}
 = \frac{1}{132.338} \;.
\end{eqnarray}
Where massive loops of top and bottom quarks are included we use
$m_t=172.5$~GeV and $m_b=4.4$~GeV. When we include the contribution from diagrams
involving a Higgs boson, the values of the width that we use are taken from
HDECAY~\cite{Djouadi:1997yw} and are shown in Table~\ref{tab:width}.

For the parton distribution functions (pdfs) we use the sets of Martin, Stirling,
Thorne and Watt~\cite{Martin:2009iq}. We use the NLO pdf fit, with
$\alpha_s(m_Z)=0.12018$ and 2-loop running. In this section we will use a common renormalization and
factorization scale equal to $m_W$. In the following sections, where the interference
with the Higgs amplitudes is studied, the common scale (unless otherwise stated) will be set equal to $m_H$.

\begin{table}[h]
\begin{center}
\begin{tabular}{|c|c||c|c|}
\hline
$m_H$~[GeV] & $\Gamma_H$~[GeV] & $m_H$~[GeV] & $\Gamma_H$~[GeV] \\
\hline
100    &   0.2573$\times 10^{-2}$   & 220    &   2.301  \\
110    &   0.2938$\times 10^{-2}$   & 240    &   3.397  \\
120    &   0.3600$\times 10^{-2}$   & 260    &    4.767 \\
130    &   0.5006$\times 10^{-2}$   & 280    &    6.443 \\
140    &   0.8281$\times 10^{-2}$   & 300    &    8.452 \\
150    &   0.1744$\times 10^{-1}$   & 320    &    10.81   \\
155    &   0.3042$\times 10^{-1}$   & 340    &   13.50
  \\
160    &   0.8255$\times 10^{-1}$   & 360    &   17.57  \\
165    &   0.2434       & 380    &   23.04   \\
170    &   0.3760       & 400    &   29.16   \\
180    &   0.6291       & 450    &   46.82   \\
190    &   1.036        & 500    &   67.94   \\
200    &   1.426        & 600    &   122.5  \\
\hline
\end{tabular}
\end{center}
\caption{Total Higgs width as a function of the Higgs mass using the HDECAY code (v3.51) 
from ref.~\cite{Djouadi:1997yw}\label{tab:width}.}
\end{table}

\subsection{Results}

We begin by assessing the numerical impact of the contribution arising from the loops containing
massive quarks, i.e. the $(t,b)$ doublet represented by the term ${\cal A}_{\rm massive}$ in Eq.~(\ref{eq:ampdecomp}).
This contribution was neglected in the analysis of
ref.~\cite{Campbell:2011bn} based on earlier studies at the 14~TeV LHC where it was found to be
very small~\cite{Binoth:2006mf}.

For now we neglect the contribution ${\cal A}_{\rm Higgs}$ in Eq.~(\ref{eq:ampdecomp})
and consider only the effect arising from adding the amplitude ${\cal A}_{\rm massive}$
to the result for two massless doublets of quarks only.
We note that adding the massive doublet results in new contributions from both ${\cal A}_{\rm massive}$
and the interference term, $2 {\rm Re} \left({\cal A}_{\rm massive}{\cal A}_{{\rm massless}}^*\right)$.
In the results that we present here we will compare three contributions to the $WW$ final state:
\vskip 0.5cm 
\begin{tabular}{r@{\hspace{0.2cm}}l}
$\sigma_{gg}[n_{\rm gen}=2]$: & the gluon-initiated contribution for two massless doublets \\ &
 resulting from the term $2{\cal A}_{\rm massless}$ in Eq.~(\ref{eq:ampdecomp}); \\ & \\
$\sigma_{gg}[n_{\rm gen}=3]$: & the gluon-initiated contribution for all three doublets \\ & 
 resulting from the combination $2{\cal A}_{\rm massless}+{\cal A}_{\rm massive}$ in Eq.~(\ref{eq:ampdecomp}); \\ & \\
$\sigma_{tot}^{NLO}$:         & the total prediction obtained by adding $\sigma_{gg}[n_{\rm gen}=3]$ and \\ &
 the NLO result ($q\bar q$, $gq$, $g\bar q$ initial states)~\cite{Campbell:1999ah,Campbell:2011bn}.
\end{tabular}
\vskip 0.5cm
Our results for these cross sections, for the production of the final state
$W^+(\to \nu_e e^+) W^- (\to \mu^- \bar{\nu}_\mu)$, are shown in Table~\ref{tab:ggWWnocuts}.
We consider various values of the hadronic collider energy $\sqrt{s}$, 
with proton-antiproton collisions at $\sqrt{s}=1.96$~TeV and proton-proton collisions otherwise.
\begin{table}
\begin{center}
\begin{tabular}{|l|c|c|c|c|c|c|}
\hline
$\sqrt{s}$~[TeV]             &1.96 ($p\bar p$)     & 7        & 8        & 10       & 12       & 14 \\ \hline 
$\sigma_{gg}[n_{\rm gen}=2]$         & 0.460(0)	& 13.74(1)  &  18.19(1) & 28.37(2) &40.06(3) &  52.99(4) \\
$\sigma_{gg}[n_{\rm gen}=3]$         & 0.490(1) 	& 15.16(1)  	 & 20.12(2) & 31.61(3)  &44.84(4)  &59.59(4)\\ 
$\sigma_{gg}[n_{\rm gen}=3]/
 \sigma_{gg}[n_{\rm gen}=2]$         & 	 1.065  &  1.103   &    1.106&  1.114&  1.119   & 1.125  \\ \hline
$\sigma_{tot}^{NLO}$         & 134.6(2)	&539(1)   & 657(1)    &  904(1)    & 1162(1)  & 1429(2) \\
$\sigma_{gg}[n_{\rm gen}=3]/
 \sigma_{tot}^{NLO}$         &	0.0036 &  0.028 &    0.030&   0.035  & 0.039    & 0.042 \\
\hline
\end{tabular}
\caption{Cross sections for the process $W^+(\to \nu_e e^+) W^- (\to \mu^- \bar{\nu}_\mu)$
(in femtobarns) with no cuts applied. The cross section resulting from the gluon-gluon initial state
($\sigma_{gg}$) is  broken down by the number of generations of quarks circulating in the loop.
 The result with three generations is also compared with the total cross section ($\sigma_{tot}^{NLO}$),
 obtained by also including the NLO corrections to the quark-antiquark initial state.
\label{tab:ggWWnocuts}}
\end{center}
\end{table}
In these calculations we have applied no cuts to the final state.

We first observe that including the third generation has only a minor effect
on the gluon-gluon initiated contribution, ranging from an increase of 6\% at the Tevatron to
an increase of 13\% at the 14~TeV LHC. This small increase is washed out even further
when considering the correction to the total cross section including $q\bar q, gq$ and $g\bar{q}$ contributions.
As expected, the gluon-gluon contribution is negligible at the Tevatron ($0.4$\% of the total) and
even at the LHC it is at most 4\% of the total at the highest energy foreseen (although this can increase upon 
application of various cuts on the final state \cite{Campbell:2011bn,Binoth:2006mf}).
We note that our findings at 14~TeV are in complete agreement with the results presented in
Ref.~\cite{Binoth:2006mf}.

\begin{figure}
\begin{center}
\includegraphics[width=8cm]{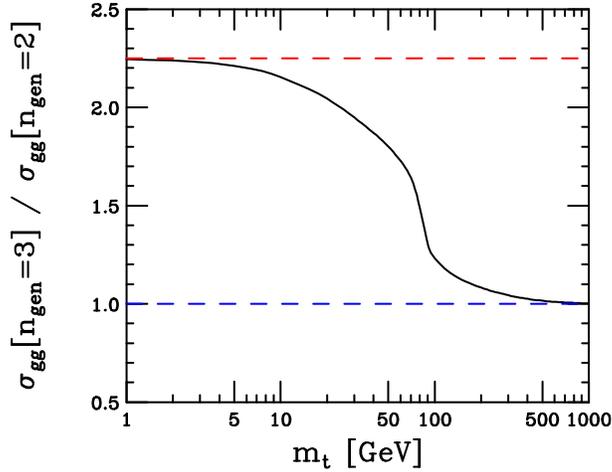}
\caption{The ratio of $\sigma_{gg}[n_{gen}=3]$ to $\sigma_{gg}[n_{gen}=2]$ as a function of the top quark mass $m_t$.
The expected results for two and three massless generations are shown as dashed blue (lower) and red (upper)
lines respectively.}
\label{fig:mtdep}
\end{center}
\end{figure}

The fact that the three-generation cross section is so close to the two-generation result may be somewhat
unexpected. For that reason, in Fig.~\ref{fig:mtdep} we illustrate the dependence of $\sigma_{gg}[n_{gen}=3]$ on
the top quark mass, $m_t$. In the limit of a very light top quark the third generation is equivalent to the
two massless generations, and as a result $\sigma_{gg}[n_{gen}=3] \to \sigma_{gg}[n_{gen}=2] \times 9/4$ as $m_t \to 0$.
In the opposite limit, when the top quark is very heavy,  the third generation decouples so that 
$\sigma_{gg}[n_{gen}=3] \to \sigma_{gg}[n_{gen}=2]$ as $m_t \to \infty$. The behaviour of the three-generation result
changes rapidly in the region $m_t \sim m_W$, as could be expected from the kinematics of the final state.

\begin{figure}
\begin{center}
\includegraphics[width=9cm]{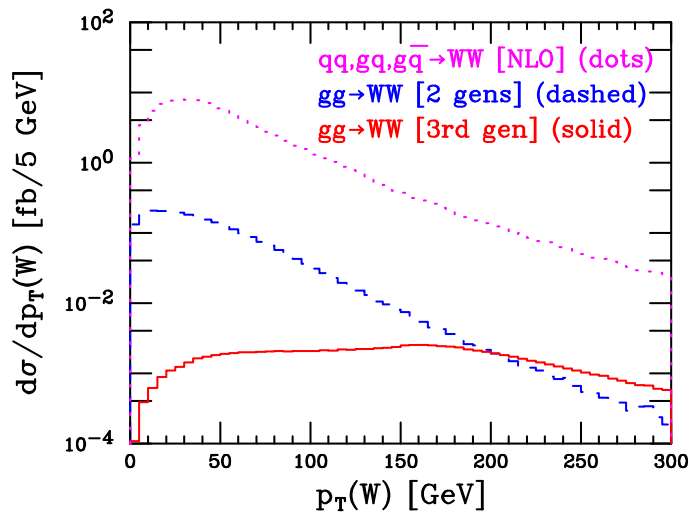}
\caption{The transverse momentum distribution of the $W$ bosons at $\sqrt s=7$~TeV and with no cuts applied.
The contribution from $gg \to WW$ through a loop of third generation quarks only is shown in solid red and the
contribution from $gg \to WW$ through two massless generations is shown in dashed blue. The distribution obtained
from the NLO calculation of the $q{\bar q},gq,g\bar{q}$ processes is shown as the dotted magenta curve.} 
\label{fig:pt_W}
\end{center}
\end{figure}

In Fig.~\ref{fig:pt_W} we present the $p_T$ distribution of one of the $W$ bosons for the first two generations (red) and
third generation only (blue). We observe that at low $p_T$ the contribution from the third generation is negligible but
in the region $p_T > m_t$ it becomes increasingly more important.
This  indicates that for searches that require a high-$p_T$ $W$ boson, the top loops should not be neglected.  Numerical
instabilities exist in the massive (and to a lesser extent in the massless)  calculation in the region of low $p^{W}_T$.
The origin of these instabilities can be traced back to terms of the form $1/\spab2.3+4.1^n$ in the $n$-point integral coefficients,
for instance in Eq.~(\ref{eq:d1coeff}) (for $n=4$). Such denominators clearly vanish in the limit that
the momentum $(3+4)$ is collinear to either particle $1$ or $2$,  i.e. the momentum $p_W=p_3+p_4$ is along the beam, so that
$p_T(W) = 0$. The amplitude may be re-expressed in terms of combinations of the scalar integrals that are finite in this limit,
a refinement that has already been partially performed for the massless amplitudes~\cite{Bern:1997sc}.
We find that some numerical instability still
remains for the massless amplitudes and, in the massive case, the denominators are both more prevalent and more cumbersome
to remove. Therefore to ensure numerical stability we instead simply impose a cut, $p^{W}_T > 2~(0.05)$~GeV in the massive
(massless) calculation. The application of this cut changes the total cross section by at most 0.05\%.

\section{Higgs $\rightarrow WW$ interference effects at the LHC}

Since we have seen that that the effect of the gluon-gluon initial state is largest at the LHC we shall
first present phenomenological results applicable for Higgs boson searches at $\sqrt{s}=7$~TeV.
In particular in this section we will consider interference effects between terms in the amplitude
that result from diagrams involving a Higgs boson and those that do not.

The cross sections that we shall compute are obtained from Eq.~(\ref{eq:ampdecomp})
by selecting various terms in the square of the amplitude, integrating over the final-state phase space and convoluting with
parton density functions appropriately. We shall divide the full contribution to the cross section as follows,
\begin{eqnarray}
{\sigma_B} &\longrightarrow& \left|{\cal A}_{\rm box}\right|^2 \;, \qquad {\cal A}_{\rm box} = 2{\cal A}_{\rm massless}+{\cal A}_{\rm massive} \;, \nn \\
{\sigma_H} &\longrightarrow& \left|{\cal A}_{\rm Higgs}\right|^2 \nn \;, \\
{\sigma_i} &\longrightarrow& 2{\rm Re} \left({\cal A}_{{\rm Higgs}} {\cal A}_{\rm box}^* \right) \;, \nn \\
{\sigma_{H,i}} &=& \sigma_H + \sigma_i \;.
\label{eq:Hidescription}
\end{eqnarray}
For a given Higgs mass the total cross section including the Higgs diagram is thus the sum of the first
three terms. $\sigma_B$ and
$\sigma_H$ are the usual leading order $gg$ background and Higgs signal cross sections respectively.
When the interference is included, we consider the quantity
$\sigma_{H,i}$ as representing the cross section for events containing a Higgs boson, to be compared
with the background-only cross section $\sigma_B$. We note that, since $\sigma_{H,i}$ is not a physical cross
section and therefore not constrained to be positive, we may (and indeed do) find that $\sigma_{H,i} < \sigma_H$.

To validate our computation of the interference terms we first compare our results with those
presented in Ref.~\cite{Binoth:2006mf} for the LHC energy of $\sqrt{s}=14$~TeV.
As already noted, in this reference the dependence on the bottom quark mass is retained but here we have
chosen to set it to zero. The expected difference, of the order of $0.1$\%, is at the same level as the
Monte Carlo uncertainty in our result. Within these uncertainties, the two calculations are in complete
agreement for a range of Higgs masses and cuts, as can be seen from Table~\ref{tab:comparebinoth}.
For the purposes of this comparison all parameters in MCFM have been
set to the values used in Ref.~\cite{Binoth:2006mf}. Ref.~\cite{Binoth:2006mf} also specifies the Higgs search cuts
used in the final three columns of the table. Although the quantity $\sigma_{H,i}$ is not explicitly given
in Ref.~\cite{Binoth:2006mf} it may easily be extracted from the results presented there.
\begin{table}
\begin{center}
\begin{tabular}{|c||c|c|c||c|c|c|}
\hline
cut selection & \multicolumn{3}{|c||}{total cross section} & \multicolumn{3}{|c|}{search cuts} \\ \hline
$M_H$ [GeV] & 140 & 170 & 200  & 140 & 170 & 200 \\ \hline
$\sigma_{H}$   & 79.90 & 116.2 & 75.39 & 1.884 & 12.95 & 1.664 \\ \hline 
$\sigma_{H,i}$ & 72.49 & 114.4 & 74.52 & 1.758 & 13.84 & 1.999 \\ \hline \hline
$\sigma_{H,i}$ (extracted from Ref.~\cite{Binoth:2006mf})
  &72.56 & 114.6 & 74.45 & 1.760 & 13.87 & 1.999 \\
\hline
\end{tabular}
\caption{A comparison of results obtained from our calculation in MCFM with the corresponding values
extracted from Table 3 of Binoth et al.~\cite{Binoth:2006mf}.
Both $W$'s decay leptonically into a single lepton flavor, no cuts are applied, $\sqrt{s}=14$~TeV and cross
sections are given in femtobarns.
The cross sections are computed either excluding ($\sigma_H$) or including ($\sigma_{H,i}$) the effect of interference
with the gluon-initiated background process.
\label{tab:comparebinoth}}
\end{center}
\end{table}

\subsection{No final state cuts} 
\label{sec:nocuts}

\begin{figure} 
\begin{center}
\includegraphics[width=12cm]{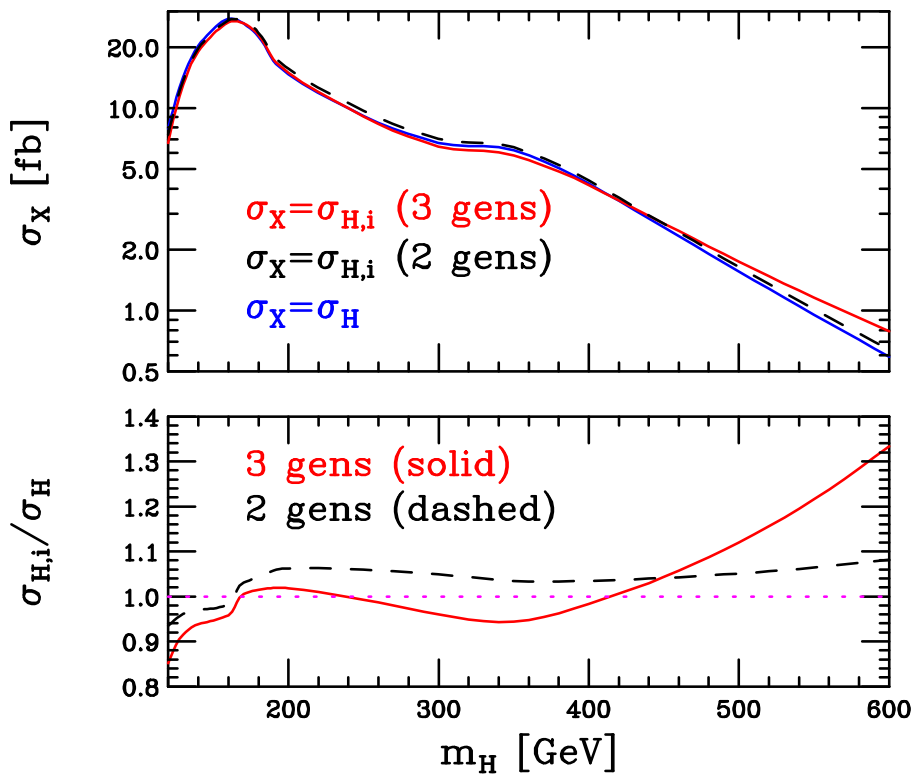} 
\caption{Upper panel: The cross sections for $gg \to H \to W^+(\to \nu_e e^+) W^- (\to \mu^- \bar{\nu}_\mu)$
in femtobarns, with ($\sigma_{H,i}$) and without ($\sigma_{H}$) the interference with SM $gg\rightarrow WW$ production. The dashed line represents the calculation
of $\sigma_{H,i}$ including only the first two generations of quarks.  Lower panel: The ratio 
of the cross sections with and without the interference terms. The dotted magenta line highlights the boundary between constructive and destructive interference.
\label{fig:higgs_bas}}
\end{center}
\end{figure}

We now return to the case at hand, namely the LHC operating at 7~TeV.
The quantities  $\sigma_H$ and $\sigma_{H,i}$, with no cuts applied on the final
state, are shown as function of $m_H$ in  Fig.~\ref{fig:higgs_bas}. Numerical
values of these cross section are shown in Table~\ref{tab:interferenceat7} for a selection
of benchmark Higgs masses.
\begin{table}
\begin{center}
\begin{tabular}{|c|c|c|c|c|c|}
\hline
$M_H$ [GeV]    & 120     & 140      & 170      & 200      & 400 \\ \hline
$\sigma_{H}$   & 7.90(1) & 20.29(1) & 26.13(2) & 14.69(1) & 4.23(1) \\ \hline 
$\sigma_{H,i}$ & 6.73(1) & 19.04(1) & 26.25(2) & 14.96(1) & 4.16(1) \\ \hline 
$\frac{\sigma_{H,i}}{\sigma_{H}}$ 
               & 0.852    & 0.938    & 1.005    & 1.018    &0.983 \\
\hline
\end{tabular}
\caption{Cross sections for $gg \to H \to W^+(\to \nu_e e^+) W^- (\to \mu^- \bar{\nu}_\mu)$ in femtobarns at $\sqrt{s}=7$~TeV with no cuts applied, computed at leading
order and either excluding ($\sigma_H$) or including ($\sigma_{H,i}$) the effect of interference
with the gluon-initiated background process.  
\label{tab:interferenceat7}}
\end{center}
\end{table}
We observe that the relative size of the interference is strongly dependent on the Higgs
mass and that the interference changes sign at the $m_H=2m_W$ threshold.
For $m_H > 2m_W$ there are two further changes of sign,
with a minimum at $m_H=2m_t$. For very large Higgs masses the interference 
becomes large and positive. For reference we have also plotted in Fig.~\ref{fig:higgs_bas} the
contribution to the interference from the first two generations of quarks only (i.e. setting
${\cal A}_{\rm massive}=0$ in the definitions of Eq.~(\ref{eq:Hidescription}).
The difference between the two and three generation results is striking. Except for values of $m_H \lesssim 200$~GeV,
in general either the sign or the magnitude of the two-generation interference is changed in the three-generation
result. This is a result of features in the three-generation result around $m_H=2m_t$, as is expected from the
thresholds of the integral functions. We conclude that, despite being a negligible factor in total rates,
it is imperative to include the massive loop contribution in calculations of interference effects. 
For the lightest Higgs mass considered here, ($m_H = 120$~GeV) the prediction for
the Higgs cross section including interference effects ($\sigma_{H,i}$) is $10-15$\% lower than
would be anticipated with the usual approach ($\sigma_H$).
Since a $\mathcal{O}(10\%)$ correction competes with the theoretical uncertainty of the
NNLO cross section these pieces should be included in the prediction
for a light Higgs cross section at the LHC. For $m_H > 400$~GeV the interference 
becomes large and constructive, although one should take care in interpreting the results
in this region since for these values of $m_H$ the Higgs width becomes very large (c.f. Table~\ref{tab:width}).
For instance, including an $\hat s$-dependent width for the Higgs boson would change the details
of the results at large $m_H$.

\begin{figure} 
\begin{center}
\includegraphics[width=10cm]{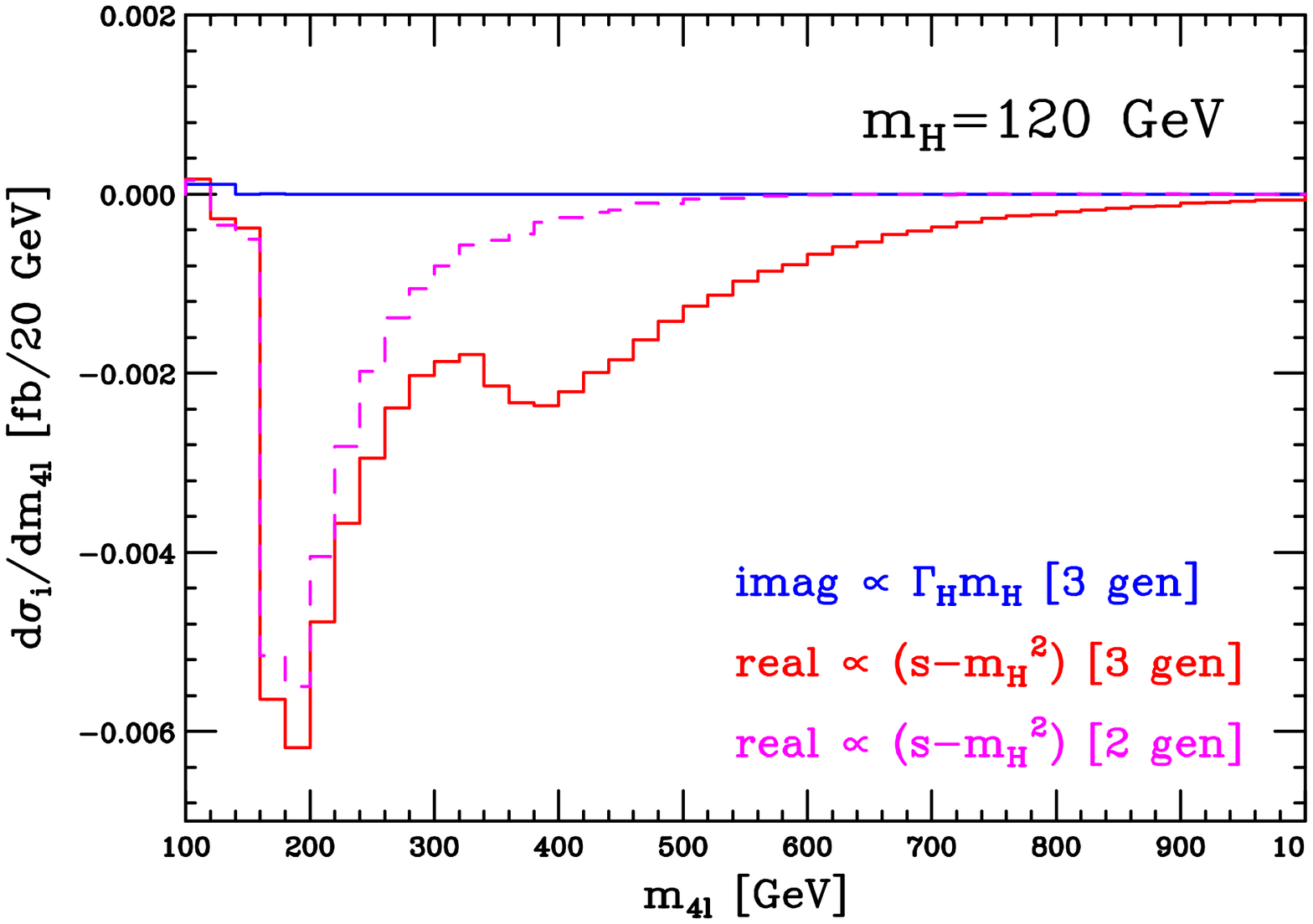} \\ \vspace*{1cm}
\includegraphics[width=10cm]{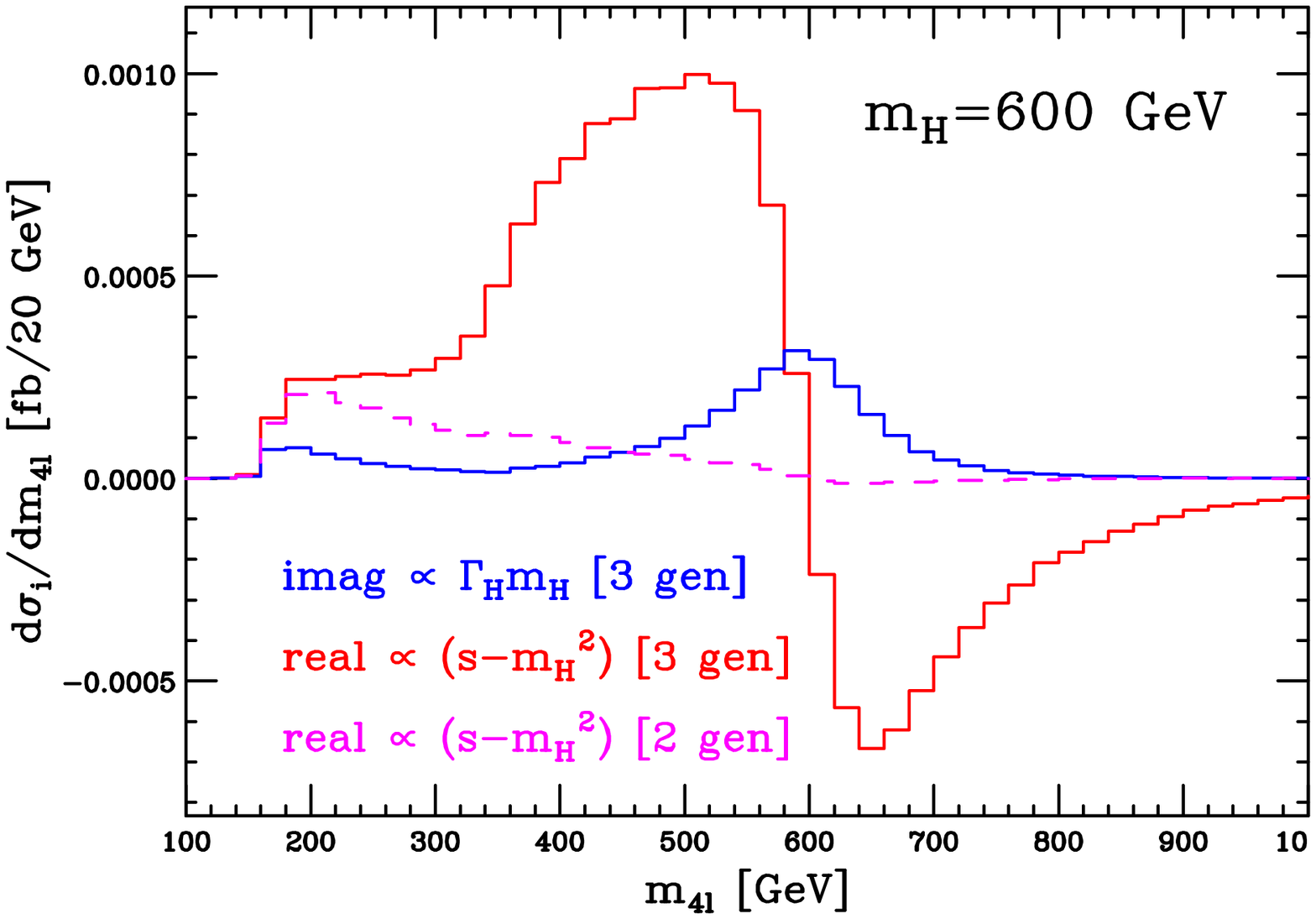} 
\caption{Contributions to the interference coming from the real (red) and imaginary (blue) part of the Higgs Breit-Wigner propagator,
as defined in Eq.~(\protect\ref{eq:intfdecomp}), for two values of the Higgs mass.
The real part of the interference with two generations of massless quarks is shown by the dashed magenta line.
\label{fig:intfdecomp}}
\end{center}
\end{figure} 
A detailed examination of the interference can be performed by separating the Breit-Wigner form of the Higgs boson propagator into
its real and imaginary parts. Thus the result for the interference can be written as,
\begin{equation}
\delta\sigma_i = \frac{(\hat s - m_H^2)}{ (\hat s - m_H^2)^2 + m_H^2 \Gamma_H^2} \, \mathfrak{Re} \left\{ 2 \widetilde{\cal A}_{\rm Higgs} {\cal A}_{\rm box}^*  \right\}
         + \frac{m_H \Gamma_H}{ (\hat s - m_H^2)^2 + m_H^2 \Gamma_H^2} \, \mathfrak{Im}  \left\{ 2 \widetilde{\cal A}_{\rm Higgs} {\cal A}_{\rm box}^*  \right\} \;,     
\label{eq:intfdecomp}
\end{equation}
where $\widetilde{\cal A}_{\rm Higgs} = \left(\hat s - m_H^2 + i m_H \Gamma_H \right) {\cal A}_{\rm Higgs}$. Although in the above equation the dependence of
$\widetilde{\cal A}_{\rm Higgs}$ and ${\cal A}_{\rm box}$ on the kinematic variables has been suppressed, both 
$\widetilde{\cal A}_{\rm Higgs}$ and ${\cal A}_{\rm box}$ contain thresholds and other kinematic structures. The results for the
two terms in Eq.~(\ref{eq:intfdecomp}) are shown in Fig.~\ref{fig:intfdecomp} as a function of 
the invariant mass of the four final state leptons, $m_{4\ell}$. Although $m_{4\ell}$ is not an experimentally-measurable quantity,
it is particularly important since it is equal to $\hat{s}$, the partonic centre of mass energy, at this
order. We show results for Higgs masses of $120$ and $600$~GeV.
For $m_H=120$~GeV the width of the Higgs boson is very small ($3.6$~MeV) and hence the contribution of the imaginary part is negligible. 
The kinematic structure in the real part means that the $\hat s$-dependence extracted in Eq.~(\ref{eq:intfdecomp}) is not a faithful representation of the full
dependence. Consequently, the dominant contribution to the interference comes from the real part of the Breit-Wigner as can be seen
in Fig.~\ref{fig:intfdecomp}. For $m_H=600$~GeV the imaginary part contributes about a third of the net interference. Although some cancellation
between the region $\hat s < m_H^2$ and $\hat s > m_H^2$ is apparent, the net contribution is positive. For reference, in Fig.~\ref{fig:intfdecomp} we
also show the contribution of the real part computed using two massless generations of quarks, which exhibits less kinematic structure.

\begin{figure} 
\begin{center}
\includegraphics[width=7cm]{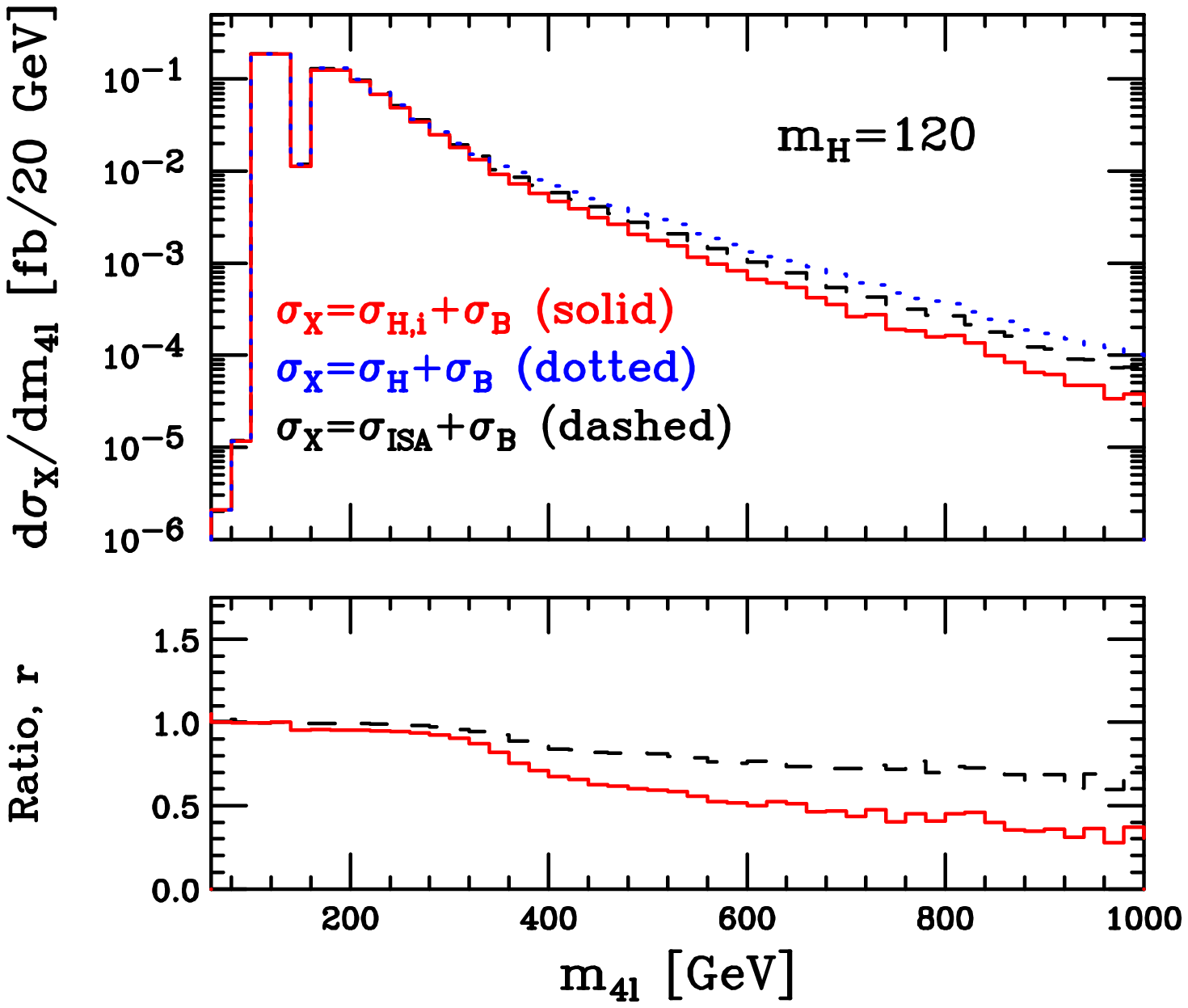}
\includegraphics[width=7cm]{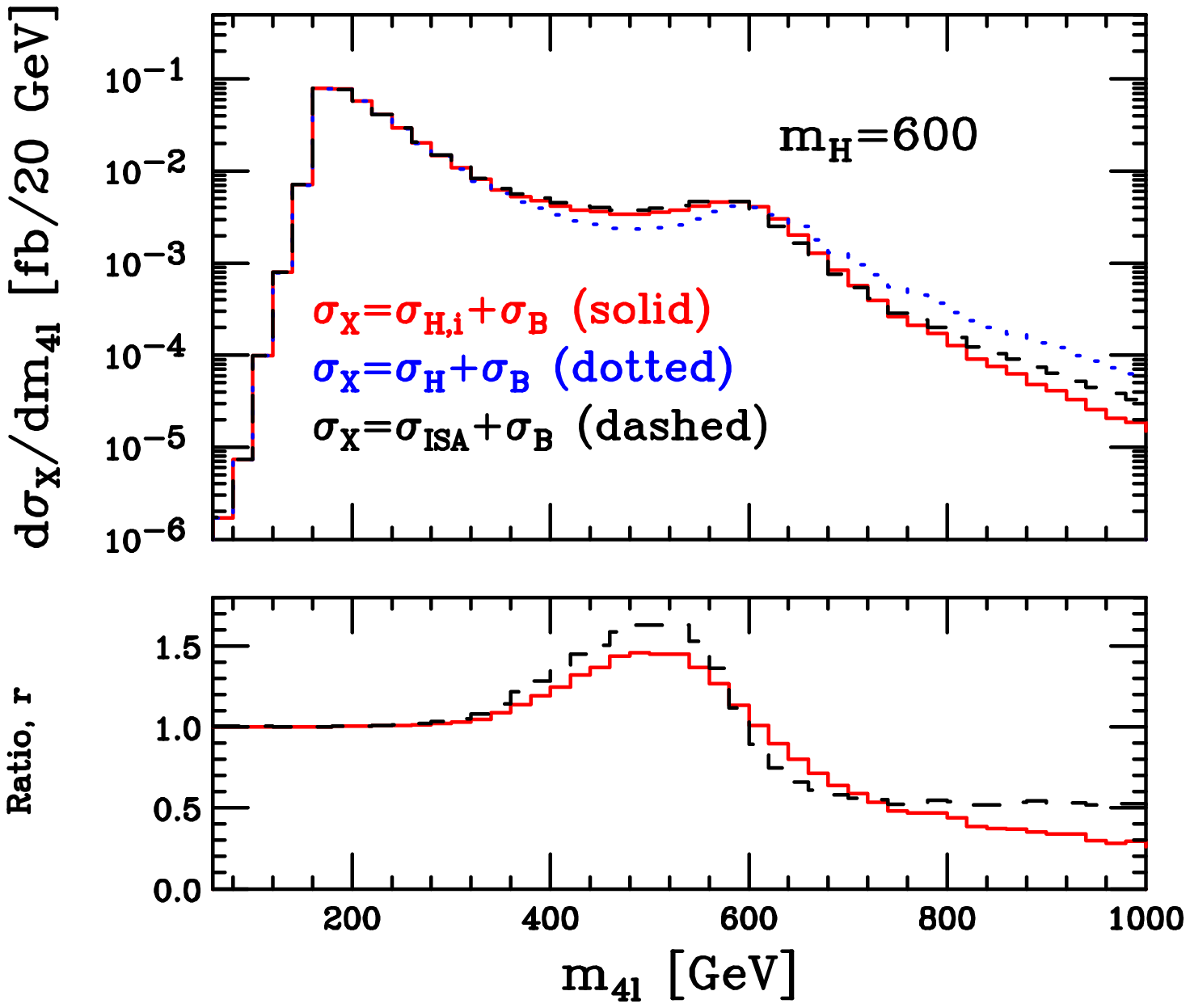} 
\caption{The upper panels present results for the invariant mass of the four final state leptons
using $\sigma_{H}+\sigma_{B}$ (blue dotted) and $\sigma_{H,i}+\sigma_{B}$ (red solid) for two different Higgs masses
(left $m_H=120$, right $m_H=600$). In addition we plot the distribution with no interference terms but using the 
rescaled Higgs propagator, $\sigma_{ISA}+\sigma_{B}$~\cite{Seymour:1995qg,Anastasiou:2011pi} (black dashed). 
In the lower panels we present the ratio
$r=\left[d(\sigma_{H,i}+\sigma_{B})/dm_{4\ell}\right]/\left[d(\sigma_{H}+\sigma_{B})/dm_{4\ell}\right]$ (red solid)
and $r=\left[d(\sigma_{ISA}+\sigma_{B})/dm_{4\ell}\right]/\left[d(\sigma_{H}+\sigma_{B})/dm_{4\ell}\right]$ (dashed) .  
\label{fig:m4lww}}
\end{center}
\end{figure} 
In Fig.~\ref{fig:m4lww} we present the distribution of $m_{4\ell}$, combining the Higgs signal ($\sigma_H$ or $\sigma_{H,i}$) with the background ($\sigma_B$), for two different Higgs masses.
We have chosen to compare the quantities $d\left(\sigma_H + \sigma_B\right)/dm_{4\ell}$
and $d\left(\sigma_{H,i} + \sigma_B\right)/dm_{4\ell}$ so that the distribution is positive throughout.  
In Refs.~\cite{Glover:1988rg,Glover:1988fe} it was shown that, for the case of $ZZ$ production, unitarity requires
destructive interference between the Higgs and non-resonant contributions in the limit of large $\hat{s}$.
From our study of the interference as a function of $m_H$, depicted in Fig.~\ref{fig:higgs_bas},
one might worry that the constructive interference at large $m_H$ violates the $\hat{s}$ requirement.
However, Fig.~\ref{fig:m4lww} illustrates that this is not the case. Although there is a net constructive interference
for $m_H=600$~GeV, this is due to the large enhancement of the cross section in the region $\hat{s} < m_H$.
In the limit of large $\hat{s}$ we observe that the interference is
large and destructive, in complete  agreement with the findings of refs.~\cite{Glover:1988rg,Glover:1988fe}.
For a much lighter Higgs boson, for instance $m_H=120$~GeV, there is no significant increase of the cross section
in the region $\hat{s} < m_H$ and the net destructive contribution simply results from the
destructive interference in the tail.

A method recently employed in ref.~\cite{Anastasiou:2011pi}  to attempt to approximate the interference effects is based
on the ``improved s-channel approximation'' (ISA) of ref.~\cite{Seymour:1995qg}. The prescription for unitarizing
the amplitude is obtained by modifying the Higgs boson propagator,
\begin{eqnarray}
\frac{i\hat s}{\hat s-m_H^2} \rightarrow \frac{im_H^2}{\hat s-m_H^2+i\Gamma_H(m_H) \frac{\hat s}{m_H}}. 
\end{eqnarray}
In order to test this approximation, in Fig.~\ref{fig:m4lww} we also show the $m_{4\ell}$ distribution calculated using this approach, denoted
$\sigma_{ISA}$. We observe that, as expected, the modified propagator decreases the cross section in the limit of large $m_{4\ell}$ ($\hat{s}$). We 
note that although the ISA approach has some of the features of the interference that we observe, there are significant differences in shape between the two approaches across the entire 
$m_{4\ell}$ range.

\subsection{Interference with search cuts} 

We now investigate the effect of the interference with more realistic
search cuts mimicking those used by the CMS and ATLAS collaborations.
To provide results for the CMS collaboration we use the cuts that were
employed in the search for the Higgs boson in the
2010 data set~\cite{Chatrchyan:2011tz}. Basic acceptance cuts
are always applied to the missing transverse energy ($\met$) and lepton rapidities,
\begin{eqnarray}
{}\met > 20 \,\,{\rm{ GeV}}, \quad |\eta_{\ell}| < 2.5 \;,
\end{eqnarray}
and then a number of further cuts are optimised
for different values of the Higgs mass. In particular, cuts
on the transverse momenta of the two leptons
($p_{T}^{\ell_{max}}$, $p_{T}^{\ell_{min}}$),
the invariant mass of the lepton pair ($m_{\ell\ell}$)
and the azimuthal angle between the leptons ($\Delta\phi_{\ell\ell}$)
are all dependent on the value of $m_H$ that is assumed in the search.
These cuts take the form,
\begin{equation}
p_T^{\ell_{max}} >  p_T^h \;,  \quad p_T^{\ell_{min}} > p_T^s \;, \quad
 m_{\ell\ell} < m_{\rm cut}  \;,  \quad \Delta\phi_{\ell\ell}  <  \Delta\phi_{\rm cut} \;,
 \label{eq:CMScuts}
\end{equation}
where the cut thresholds for some benchmark values of
$m_H$ are presented in Table~\ref{table:cmscuts}.
\begin{table}
\begin{center}
\begin{tabular}{|c|c|c|c|c|}
\hline
$m_H$ [GeV]    & $p_T^h$~[GeV] & $p_T^s$~[GeV]  & $m_{\rm cut}$~[GeV] & $\Delta\phi_{\rm cut}$ \\ \hline
130            & 25      & 20	    & 45	     & $60^\circ$  \\ \hline 
160            & 30      & 25	    & 50 	     & $60^\circ$  \\ \hline 
200            & 40      & 25	    & 90 	     & $100^\circ$ \\ \hline 
\end{tabular}
\caption{Kinematic cuts used by CMS (taken from Ref.~\cite{Chatrchyan:2011tz}) for a selection of potential Higgs masses. Cuts represent 
the lower limits on the lepton $p_{T}$ for the hardest and softest leptons, the upper limit on the invariant mass of the leptons $m_{\ell\ell}$ and
the maximum azimuthal angular separation between the leptons, $\Delta\phi_{\ell\ell}$ (c.f. Eq.~(\protect\ref{eq:CMScuts})).
\label{table:cmscuts}}
\end{center}
\end{table}

The cuts adopted by the ATLAS collaboration~\cite{Aad:2011qi} are as follows.
The set of basic acceptance cuts is,
\begin{equation} 
p_T^{\ell_{max}} > 20 \,\, {\rm{GeV}}, \quad p_T^{\ell_{min}} > 15 \,\, {\rm{GeV}}, \quad
\met > 30 \,\,{\rm{ GeV}}, \quad |\eta_{\ell}| < 2.5 \;,
\label{eq:ATLAScuts}
\end{equation}
and a further two cuts are applied that depend only on whether or not the putative
Higgs boson is lighter than $170$~GeV,
\begin{equation} 
m_{\ell\ell} < 50\,\, (60)\,\,  {\rm{GeV}}, \quad \Delta\phi_{\ell\ell} < 1.3 \,\,(1.8) \;,
\label{eq:ATLASmhdep}
\end{equation}
(numbers in parentheses indicate the values for $m_H>170$~GeV).
In addition, the transverse mass $M_T$ is constrained to be in the region $0.75 \, \,m_{H}  < M_{T} < m_{H}$, where $M_T$ is defined as,
\begin{eqnarray} 
M_T=\sqrt{(E^{\ell\ell}_T+E_{T}^{\rm{miss}})^2-({\bf{p}}^{\ell\ell}_T+{\bf{p}}_T^{\rm{miss}})^2}
\end{eqnarray}
and $E^{\ell\ell}_{T}=\sqrt{({\bf{p}}_{T}^{\ell\ell})^2+m_{\ell\ell}^2}$.

\begin{figure} 
\begin{center}
\includegraphics[width=7.5cm]{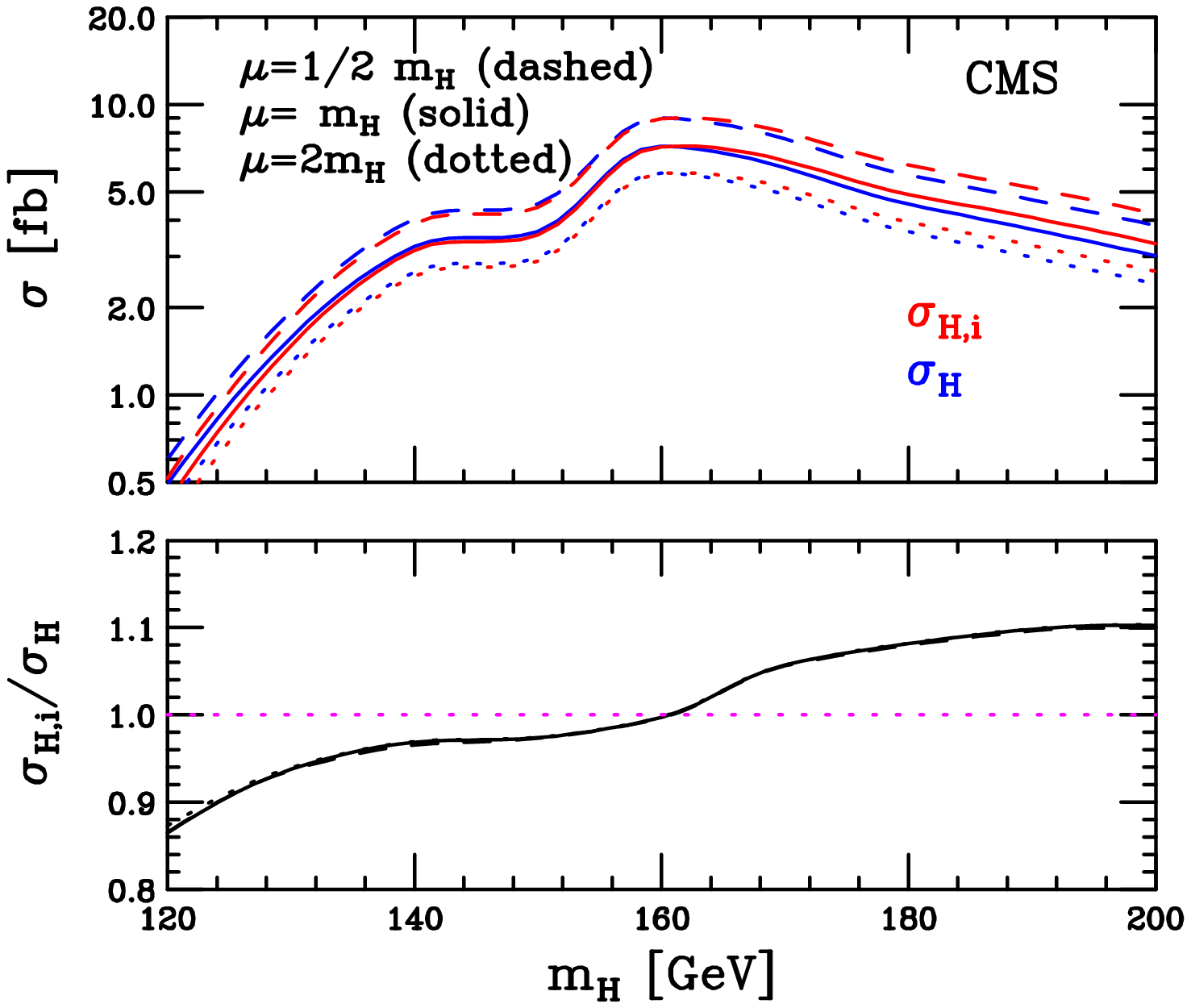} \includegraphics[width=7.5cm]{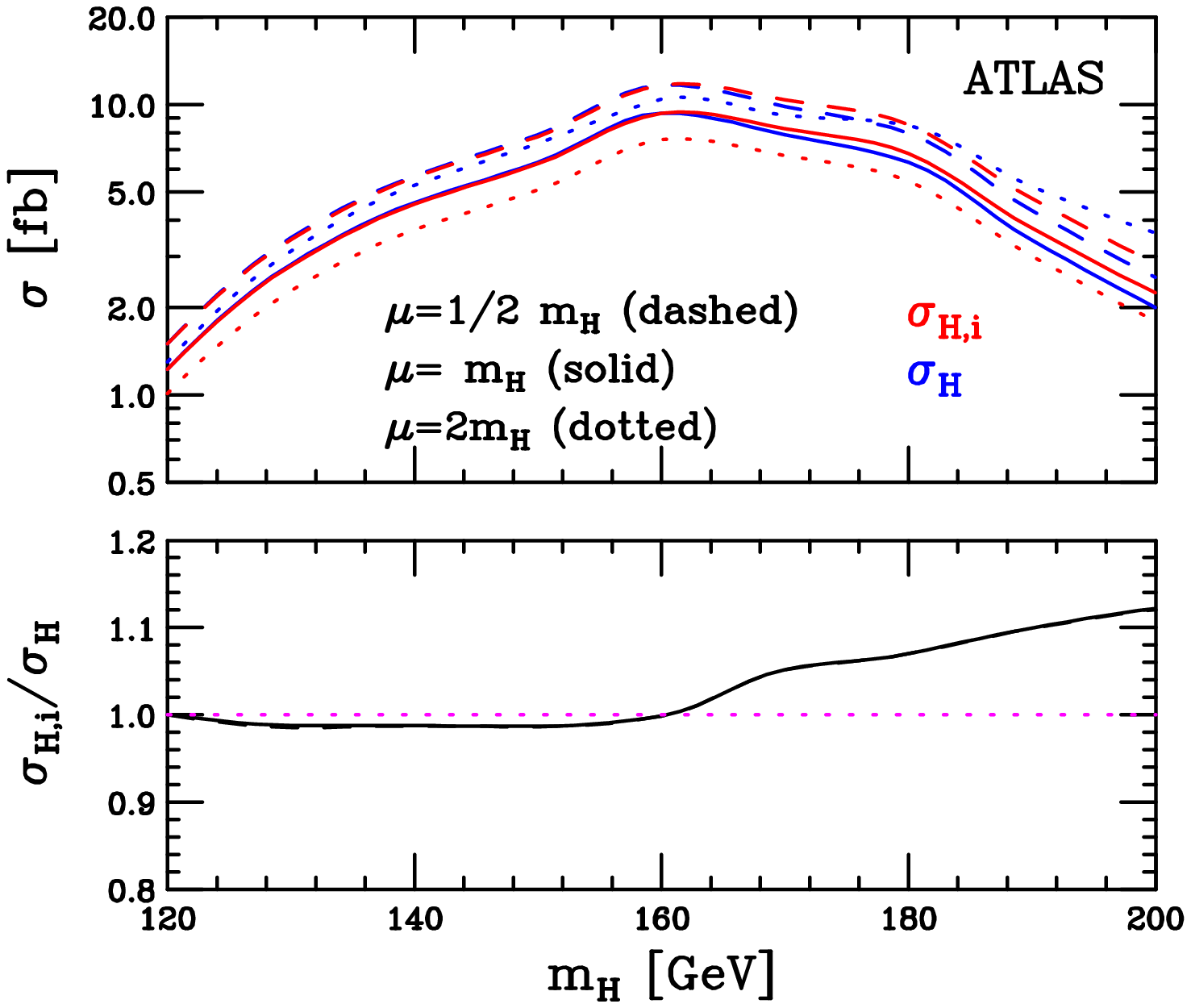} 
\caption{Cross sections for $gg \to H \to W^+(\to \nu_e e^+) W^- (\to \mu^- \bar{\nu}_\mu)$ in femtobarns with and without interference effects using
the CMS (left) and ATLAS (right) Higgs search cuts \cite{Chatrchyan:2011tz,Aad:2011qi} The red curves
include the effects of the interference whilst the blue curves do
not. We evaluate the cross section at three different scales $m_{H}/2$
(dashed), $m_H$ (solid) and $2m_H$ (dotted). The lower panel shows the ratio between the cross sections with and without
the interference, evaluated using each of the three scale choices.
\label{fig:higgs_search} }
\end{center}
\end{figure}

Our results for the cross sections with each set of cuts, with and without the interference terms, are
depicted in Fig.~\ref{fig:higgs_search} and tabulated for three values of $m_H$ in Table~\ref{table:cmsh}.
\begin{table}
\begin{center}
\begin{tabular}{|c||c|c|c||c|c|c|}
\hline
Experiment & \multicolumn{3}{|c||}{CMS} & \multicolumn{3}{|c|}{ATLAS} \\ \hline
$M_H$ [GeV] & 130 & 160 & 200  & 130 & 160 & 200 \\ \hline
$\sigma_{H}$   &   1.58 &		7.18 &  	3.02	& 2.82 &	9.35  & 2.00 \\ \hline 
$\sigma_{H,i}$ & 	1.48 &	7.16 & 3.33	 &	2.79	  &9.34	  & 2.24 \\ \hline 
$\frac{\sigma_{H,i}}{\sigma_{H}}$   &0.94  &  1.00 &1.10  & 0.99 & 1.00 & 1.12 \\\hline
\end{tabular}
\caption{Cross sections for $gg \to H \to W^+(\to \nu_e e^+) W^- (\to \mu^- \bar{\nu}_\mu)$ in femtobarns at $\sqrt{s}=7$~TeV with typical experimental Higgs search cuts
applied (see Ref.~\cite{Chatrchyan:2011tz,Aad:2011qi}), computed at leading
order and either excluding ($\sigma_H$) or including ($\sigma_{H,i}$) the effect of interference
with the gluon-initiated background process. 
\label{table:cmsh}}
\end{center}
\end{table}
For this study we have focused primarily on lighter Higgs bosons, where the $H \to WW$ search channel is most relevant,
and have considered three different choices of scale, $(m_H/2, m_H, 2m_H)$ when evaluating our predictions.
At first sight the results appear qualitatively similar. As might be expected, the relative effect of the interference
is insensitive to the choice of scale and grows with $m_H$, as already observed for the case of no cuts in Fig.~\ref{fig:higgs_bas}.
Although the ratio of $\sigma_{H,i}$ to $\sigma_{H}$ is similar for both sets of cuts
above $160$~GeV, there are major differences in the overall impact of the interference terms in the low mass region, $m_{H} \lesssim 140$ GeV.
In this region one finds that, using the CMS cuts, the
interference is destructive and leads to a reduction of around $10\%$
in the Higgs signal cross section. On the other hand, the
result of the interference terms after the application of the ATLAS
cuts is an order of magnitude smaller, around 1\%.

\begin{figure} 
\begin{center}
\includegraphics[width=9cm]{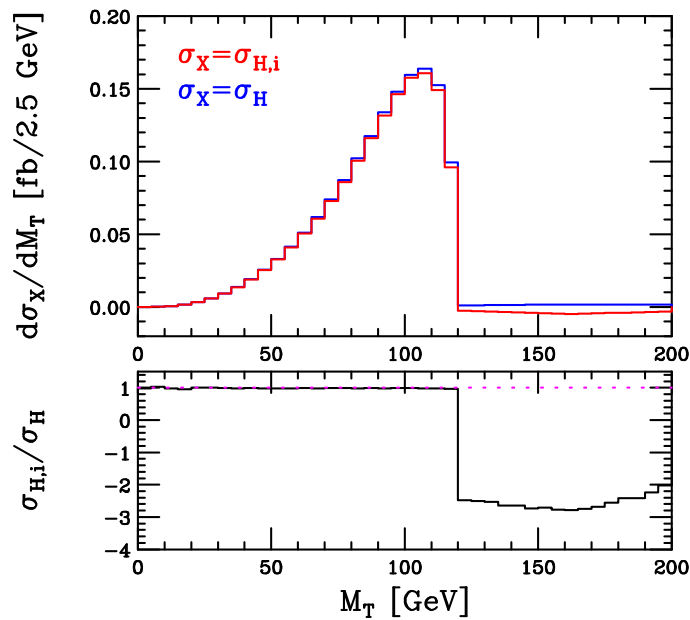}  
\caption{The $M_T$ distribution for a Higgs mass of 120 GeV with no cuts applied (upper panel). The lower (red) histogram
is calculated including the effects of the interference term whilst the upper (blue) curve is obtained using the Higgs
amplitude only. The ratio of the red and blue curves is shown in the lower panel.
\label{fig:mt}}
\end{center}
\end{figure}

The large destructive interference effects under the CMS cuts 
are not surprising since similar results are found when no cuts are
applied, c.f. Fig.~\ref{fig:higgs_bas}. We now investigate
the ATLAS cuts in order to understand the origin of these destructive effects
in the low mass region and why they are absent for this choice of cuts. 
The only major difference
between the ATLAS and the CMS (or no) cuts is the application of the
cut on the transverse mass of the system. We therefore present the $M_T$
distribution for $m_H=120$ GeV, with no cuts applied, in
Fig.~\ref{fig:mt}.
It is clear that the $M_T$ distribution has a kinematic edge at $M_T=m_H$ since,
if the Higgs were always produced on-shell, the transverse mass
$M_T$ would be zero at this order in the region $M_T > m_H$. However, small finite width effects
ensure that this distribution has a small (but non-zero) tail. This argument holds for the resonant Higgs contributions
to the amplitudes, but the non-resonant box diagrams have no such kinematic edge and instead can be relatively large in
the $M_T > m_H$ region. As a result the interference terms in this
region have a large effect.
In the region $M_T > m_H$ we see that the prediction for what appears to be a
cross section is actually negative. 
However $\sigma_{H,i}$ is not a true physical cross section, it is merely the
change in the cross section induced by the presence of the Higgs boson in the
calculation of the relevant amplitudes. To obtain a physical cross section, one
would have to add the contribution from the background squared,
$\sigma_{B}$. This contribution is large and positive and as a result
the physical cross section remains positive. The destructive interference at large $M_T$ is of course
just a reflection of the expected destructive interference at large $\hat s$ already discussed in Section~\ref{sec:nocuts}
and illustrated in Figs.~\ref{fig:intfdecomp} and~\ref{fig:m4lww}. 

\begin{figure} 
\begin{center}
\includegraphics[width=7.5cm]{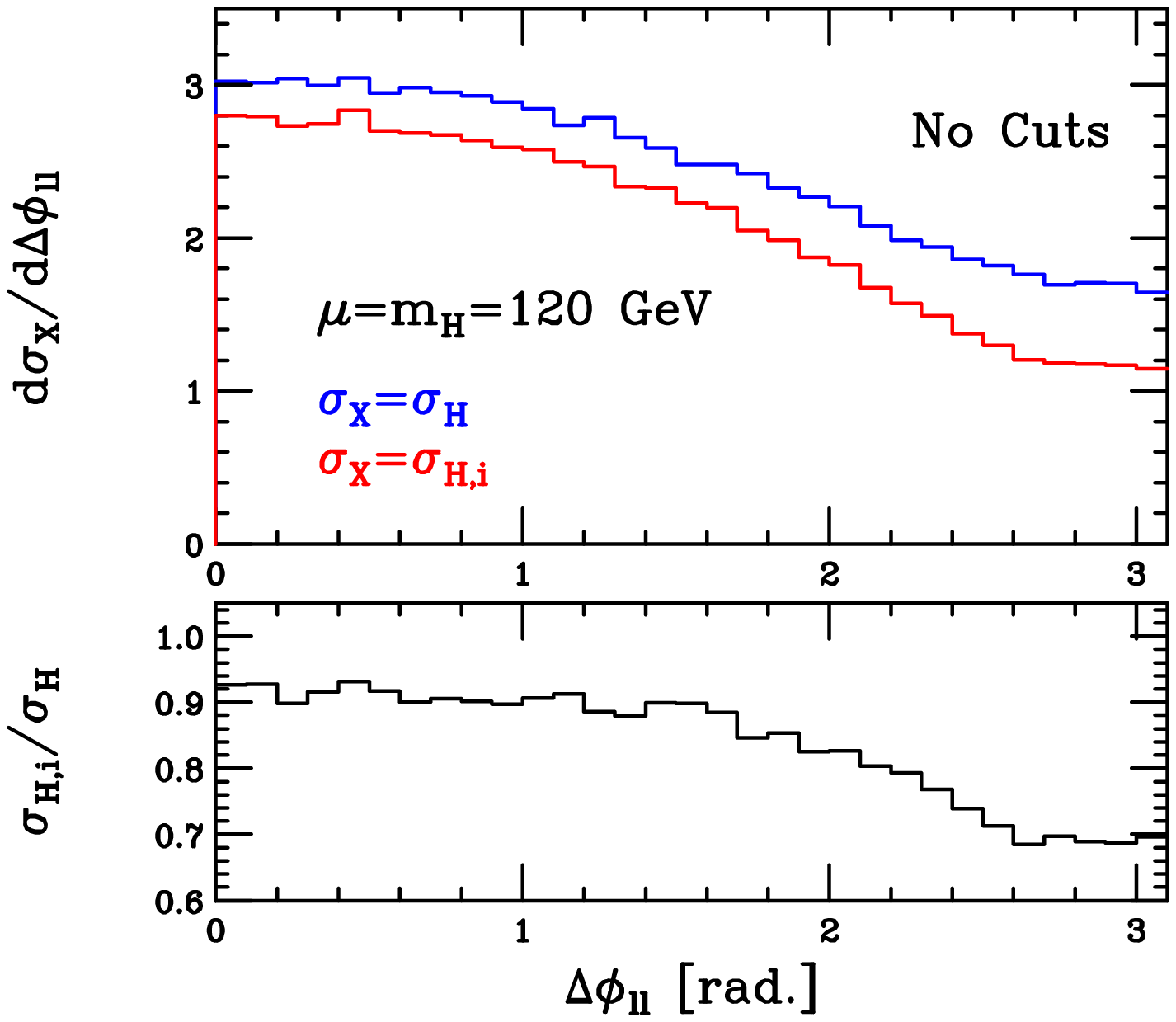} 
\includegraphics[width=7.5cm]{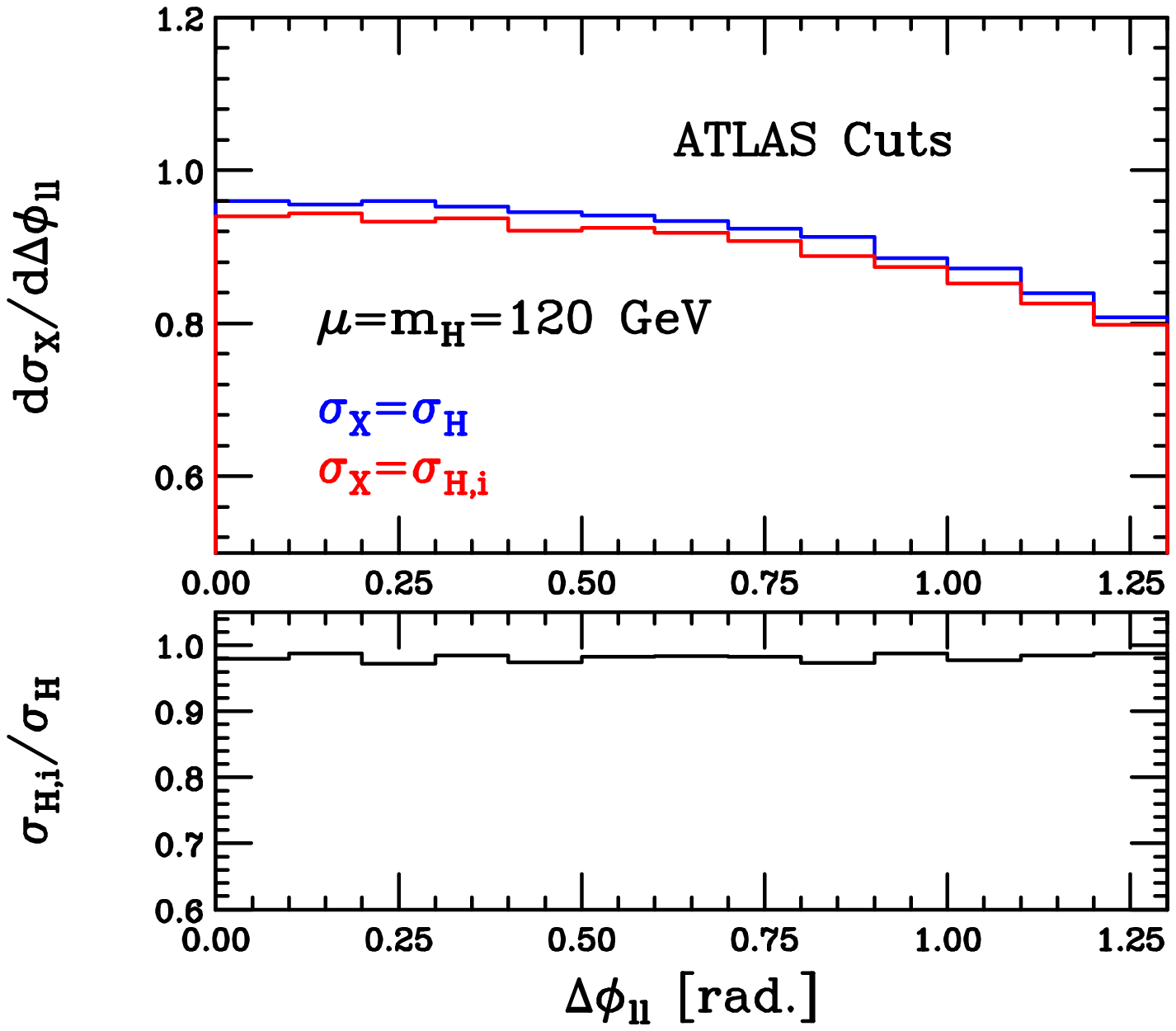} 
\caption{The $\Delta\phi_{\ell\ell}$ distribution for a Higgs mass of 120 GeV, both with no cuts applied (left)
and using the ATLAS cuts (right). The lower (red) histograms are calculated including the effects of the 
interference ($\sigma_{H,i}$) while the upper (blue) curves are obtained using the Higgs amplitudes ($\sigma_H$) only.
In the lower panel we plot the ratio of the two curves.
\label{fig:phi}}
\end{center}
\end{figure}

From this figure we can therefore conclude that the $M_T$ cut employed by ATLAS naturally removes the region of
phase space in which the interference is destructive and, as a result
$\sigma_{H,i}/\sigma_{H} \sim 1$ under the ATLAS cuts. It is also
important, however, to quantify the effect of the interference on
other kinematic distributions. One important example is the
azimuthal angle between the leptons ($\Delta\phi_{\ell\ell}$) which is
used in the search cuts of Eqs.~(\ref{eq:CMScuts},\ref{eq:ATLASmhdep}).
Since the Higgs signal peaks at low $\Delta\phi_{\ell\ell}$ and the 
background peaks in the region close to $\Delta\phi_{\ell\ell} = \pi$, this variable is a strong discriminant between
signal-like and background-like events.
Our predictions for this distribution, using $m_H=120$~GeV, are shown in Fig.~\ref{fig:phi}.
We observe that with no cuts applied there is a non-trivial shape change in $\Delta\phi_{\ell\ell}$.
However, after the application of the cuts in Eqs.~(\ref{eq:ATLAScuts},\ref{eq:ATLASmhdep}) and the $M_T$ constraint,
the shape change is negligible. This suggests that it is safe to use the ratio $\sigma_{H,i}/\sigma_{H}$ as an overall re-weighting
constant under these cuts.

To summarise, we would advocate the use of an upper limit on $M_T$ in Higgs searches in the $WW$ channel,
for instance as currently used by the ATLAS collaboration.
In addition to serving as a cut that reduces Standard Model backgrounds, it also reduces the
destructive interference between the SM and Higgs amplitudes by around 10\%, thereby increasing the expected
Higgs signal.

\section{Higgs $\rightarrow WW$ interference effects at the Tevatron}

The CDF and D0 experiments at the Tevatron also use the $WW$ channel 
to search for the Higgs boson in the mass range
$120-200$~GeV~\cite{Aaltonen:2011gs}.
Since we are discussing gluon-gluon initial states
the difference between the $p\bar p$ and $pp$ colliders is immaterial and 
for this analysis the Tevatron and the LHC differ only in their centre of mass energies.
The Higgs cross section at the Tevatron has been studied in great
detail~\cite{Anastasiou:2002yz,Ravindran:2003um,Anastasiou:2004xq,Anastasiou:2007mz,Catani:2007vq,Grazzini:2008tf,Anastasiou:2011pi},
resulting in a theoretical uncertainty that is taken to be ${\cal O}(10\%)$~\cite{Aaltonen:2011gs}.
Since the interference effects that we have studied so far at the LHC contribute at precisely this level, in this section
we perform a more detailed study in order to quantify the effects in the Tevatron search region.

We examine the effect of the interference both with no cuts and with typical search cuts used in the experimental analyses.
Since the cuts used by CDF and D0 do not differ significantly, we choose here
to focus on the set of cuts employed in recent 
CDF analyses~\cite{Aaltonen:2011gs,CDFnote9887}. Specifically, we require, 
\begin{eqnarray} 
 && p_{T}^{\ell_{max}} > 20~{\rm {GeV}}, \quad |\eta^{\ell_{max}}| < 0.8, \nn \\
 && p_{T}^{\ell_{min}} > 10~{\rm {GeV}}, \quad |\eta^{\ell_{min}}| < 1.1, \quad m_{\ell\ell} > 16~{\rm {GeV}} \;.
\end{eqnarray}
In addition the missing transverse momentum is constrained using the $\met^{\rm spec}$  variable defined by~\cite{CDFnote9887},
\begin{equation}
\met^{\rm spec} = \met \sin \left[ {\rm min} \left( \Delta \phi , \, \frac{\pi}{2} \right) \right] \;.
\end{equation}
$\Delta \phi$ is the distance between the $\met$ vector and the nearest lepton or jet.
We require that $\met^{\rm spec} > 25 \;{\rm GeV}$.

\begin{figure} 
\begin{center}
\includegraphics[width=12cm]{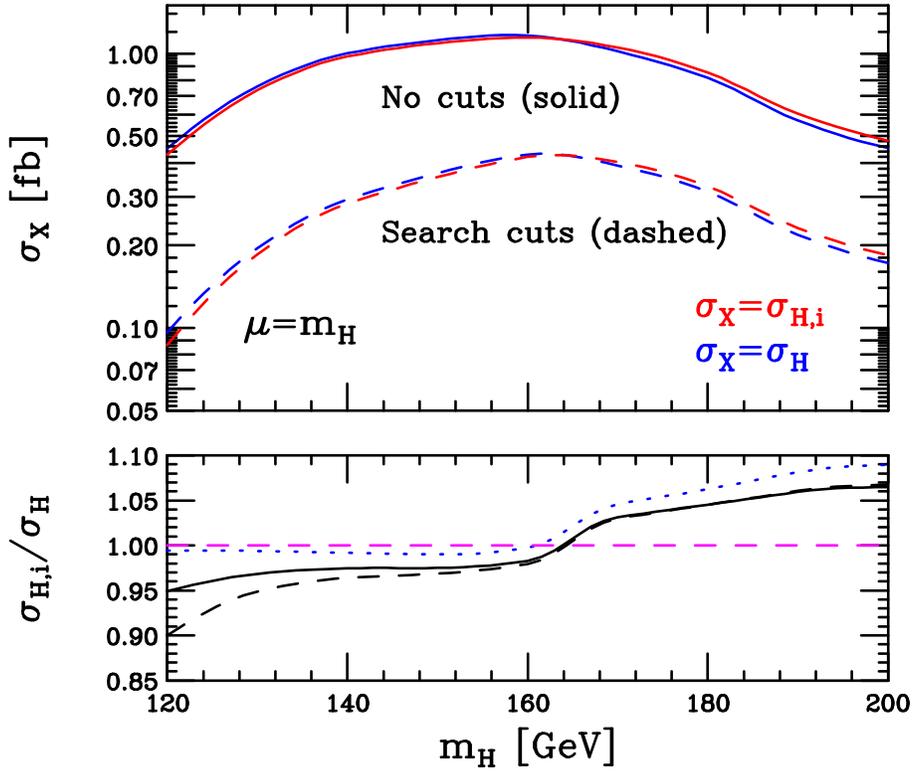} 
\caption{Cross sections for $gg \to H \to W^+(\to \nu_e e^+) W^- (\to \mu^- \bar{\nu}_\mu)$ in femtobarns at the Tevatron,
with and without interference effects. The cross sections are computed
by applying the Higgs search cuts (lower curves) or not (upper curves). The red curves include the effects of the interference 
whilst the blue curves do not. The lower panel shows the ratio $\sigma_{H,i}/\sigma_H$ for no cuts (solid), search
cuts (dashed) and search cuts with an additional constraint $M_T < m_H$ (dotted).\label{fig:tev}}
\end{center}
\end{figure}

Our results both with and without this set of cuts are presented in Fig.~\ref{fig:tev}.
We observe that with no cuts the interference has the same structure
as at the LHC. For $m_H < 2m_W$ the effect is destructive and of the order
5\% for $m_H < 130$~GeV. Around $m_H \sim 2m_W$ the effects are very
small whilst for higher $m_H$ the effect is around 5\% and
constructive.  We also observe that applying the search cuts
increases the magnitude of the destructive interference in the low
mass region. This effect, at the level of $10$\% for $m_H=120$~GeV, is as large
as the theoretical uncertainty estimated from the NNLO cross section. The effect of the
interference on current Tevatron analyses should therefore be taken into account.

In the previous section we observed that a cut on the transverse mass $M_T$ effectively
eliminated the region of destructive interference.
We have therefore recalculated the quantities $\sigma_{H,i}$ and $\sigma_{H}$ at the Tevatron, using
the same CDF cuts described above but with an additional requirement that
$M_T < m_H$. Under these cuts we indeed find that the impact of the interference is reduced
considerably, from $\mathcal{O}(10\%)$ to $\mathcal{O}(1\%)$ just as at the LHC.
The results for the interference with the additional $M_T$ cut are 
illustrated by the dotted blue curve in Fig.~\ref{fig:tev}.

\section{Combining the interference with NNLO predictions}

In this paper we have computed the interference between Higgs and continuum contributions at leading order,
i.e. between the production of a Higgs boson via a top (or bottom) quark triangle and a 
$gg\rightarrow WW$ box diagram. To calculate the interference at the next order 
one would need the NLO corrections to the Higgs process (which are known) and the NLO corrections to the gluon-gluon process 
(which are not). The NNLO result for the Higgs process has of course been known for some
time~\cite{Anastasiou:2002yz,Ravindran:2003um,Anastasiou:2004xq,Anastasiou:2007mz,Catani:2007vq,Grazzini:2008tf,Anastasiou:2011pi},
with large (factor of two) corrections to the LO cross section.

A natural question arises as to how our results for the interference should be combined 
with the NNLO Higgs production cross section.  One could simply modify the NNLO cross section by adding the LO
interference terms, i.e. by defining,
\begin{eqnarray} 
\sigma^{NNLO}_{H,i} = \sigma^{NNLO}_{H} + (\sigma^{LO}_{H,i} -\sigma^{LO}_{H}) \;. 
\label{eq:int1}
\end{eqnarray} 
Since $\sigma^{NNLO}_{H}$ is much larger than $\sigma^{LO}_{H,i}$ one would expect that using Eq.~(\ref{eq:int1})
to estimate the effect of the interference would reduce the results in this paper by approximately a factor of two. 

However, treating the interference as an absolute correction to the NNLO cross section neglects the
effect of all higher orders on the interference. Although the Higgs process receives large corrections at NLO and
NNLO, the corresponding corrections to the gluon-gluon box diagram processes are unknown.
Since we are not able to quantify these corrections a more conservative 
approach (particularly in the low-mass region where interference effects are large) would be to re-weight
the NNLO cross section by the relative effect of the interference at LO, i.e. define,
\begin{eqnarray} 
\sigma^{NNLO}_{H,i} = \sigma^{NNLO}_{H}\bigg(\frac{\sigma_{H,i}^{LO}}{\sigma_{H}^{LO}}\bigg).  
\label{eq:int2}
\end{eqnarray} 
It is this approach that we recommend in the region where the interference is destructive.

\section{Conclusions}

In this paper we studied the production of $WW$ pairs through gluon
fusion in the Standard Model. This process proceeds at loop-level through a closed
fermion loop. In a previous paper~\cite{Campbell:2011bn} we presented
analytic results for massless fermions, which correspond to the first
two generations circulating in the loop.  We extended these results to
include contributions from the third generation keeping the
full dependence on the top mass $m_t$. These formulae have been
implemented into MCFMv6.1 which will soon be publicly available. The
process $gg\rightarrow WW \rightarrow \nu \ell^+ \ell'^{-} \bar{\nu}'$
keeping the full top (and bottom) mass dependence was previously
calculated in ref.~\cite{Binoth:2006mf} using a semi-numerical approach.

The amplitude for $gg\rightarrow WW$ with non-zero $m_t$ can be
written in the usual one-loop expansion in terms of scalar integrals.
We found that the rational terms and all but two bubble coefficients
were identical to the massless result. Further, the remaining two
bubble coefficients were found to sum to the equivalent $m_t=0$
coefficient. As a result we only needed to calculate one bubble
coefficient. We used generalised unitarity techniques to calculate the
box and triangle coefficients, which differ from the massless results.

One of the most important phenomenological applications of these
results is the effect on the Higgs/SM continuum interference. We
illustrated that in general there are big deviations between the 2 and 3 
generation calculations of this interference. It is crucial to
quantify the impact of the interference as a function of the putative
Higgs mass and under a variety of cuts, since in general these terms
can contribute changes in rates comparable to the theoretical
uncertainty associated with a NNLO prediction for the cross section.
We found that in general the interference is largest far away from the
$2m_W$ threshold.  For large $m_H$ $( > 2m_t) $ the interference is
constructive and for low ($m_H < 140$~GeV) masses it is destructive.  In
the region $120 < m_H < 130$~GeV the destructive interference reduces
the total cross section by around $10-15$\%. This large interference
originates from non-resonant contributions above $m_H$.

We illustrated that the interference is sensitive to the experimental
cuts under consideration and that cuts used in the 2010 ATLAS
analysis~\cite{Aad:2011qi} dramatically reduce the large destructive
interference for small $m_H$. We showed that this is a result of the
$M_T$ cut that is employed, in particular the upper bound that effectively
removes the non-resonant contribution to the interference.
After applying the cut $M_T < m_H$, the interference is
reduced from $\mathcal{O}(10\%)$ to
$\mathcal{O}(1\%)$. Therefore we would advocate the use of the cut
$M_T < m_H$ for all experiments at hadron colliders since this
enhances the signal by removing the region in which the destructive
interference dominates.

The total $gg\rightarrow H$ cross section is known to 
NNLO~\cite{Anastasiou:2002yz,Ravindran:2003um,Anastasiou:2004xq,Anastasiou:2007mz,Catani:2007vq,Grazzini:2008tf,Anastasiou:2011pi}
accuracy and the $K$-factor from going from LO to NNLO is
large. Therefore, if the interference is treated as an absolute
correction to the NNLO cross section its impact is reduced by an
approximate factor of two. However, the interference is present at
every order in perturbation theory and as such a more conservative
approach would be to re-weight the NNLO cross section by the
relative effect of the interference at LO. The size of the interference
means that this is particularly important when setting Higgs mass limits
using this channel for $m_H \lesssim 140$~GeV.

\section*{Acknowledgements} 

We thank Joey Huston, Bruce Mellado, Kirill Melnikov and Frank Petriello for useful discussions.
Fermilab is operated by Fermi Research Alliance, LLC under Contract No. DE-AC02-07CH11359 with the
United States Department of Energy.

\bibliography{CEWfinal}
\bibliographystyle{JHEP}

\end{document}